\documentclass[traditabstract]{aa}
\usepackage{amsmath}
\usepackage{amssymb}
\usepackage{booktabs}
\usepackage{multirow}
\usepackage{nicefrac}
\usepackage{txfonts}
\usepackage{epsfig,graphicx}
\usepackage{units}
\usepackage{bbold}
\usepackage[colorlinks=true, linkcolor=blue, citecolor=blue, urlcolor=blue]{hyperref}
\usepackage{natbib}
\bibpunct{(}{)}{;}{a}{}{,}  
\newcommand{\txd}{\text{d}}
\newcommand{\cext}{C_\text{ext}}
\newcommand{\cpol}{C_\text{pol}}
\newcommand{\ccpol}{C_\text{cpol}}
\newcommand{\csca}{C_\text{sca}}
\newcommand{\cscapol}{C_\text{sca,pol}}
\newcommand{\bfk}{{\boldsymbol{k}}}

\newcommand{\bfo}{{\boldsymbol{o}}}

\newcommand{\bfK}{{\boldsymbol{K}}}
\newcommand{\bfZ}{{\boldsymbol{Z}}}
\newcommand{\bfr}{{\boldsymbol{r}}}
\newcommand{\bfR}{{\boldsymbol{R}}}

\newcommand{\bfE}{{\boldsymbol{E}}}

\newcommand{\bfM}{{\boldsymbol{M}}}
\newcommand{\bfn}{{\boldsymbol{n}}}

\newcommand{\bfd}{{\boldsymbol{d}}}
\newcommand{\E}{\text{E}}
\newcommand{\N}{\text{N}}
\newcommand{\bfS}{{\boldsymbol{S}}}

\newcommand{\e}[1]{\textup{e}^{#1}}

\newcommand{\degree}{{\ensuremath{^\circ}}}

\newcommand{\simgreat} {\mathbin{\lower 3pt\hbox{$\rlap{\raise
        5pt\hbox{$\char'076$}}\mathchar"7218$}}}

\newcommand{\simless}{\mathbin{\lower 3pt\hbox {$\rlap{\raise
        5pt\hbox{$\char'074$}}\mathchar"7218$}}}

\newcommand{\name}{{\sc{MCpol}}}

\usepackage[normalem]{ulem}                                       %
\usepackage{color}                                                %
\definecolor{new}{RGB}{40, 160, 40}                               %
\definecolor{com}{RGB}{240, 040, 040}                             %
\definecolor{rem}{RGB}{100, 100, 100}                             %
\defcitealias{Peest2017}{P17}

\begin{document}

\title{Paradigmatic examples for testing models of optical light
  polarization by spheroidal dust}

\titlerunning{Testing models for polarization by spheroids in opaque objects}

\author{C. Peest\inst{1,2} \and
  R. Siebenmorgen\thanks{contact: Ralf.Siebenmorgen@eso.org}\inst{1} \and
  F. Heymann\inst{3} \and T. Vannieuwenhuyse\inst{1} \and M. Baes\inst{2}}

\institute{{European Southern Observatory, Karl-Schwarzschild-Str. 2,
    D-85748 Garching b. M\"unchen, Germany}
  \and {Sterrenkundig Observatorium, Universiteit Gent, Krijgslaan 281
    S9, 9000 Gent, Belgium}
  \and {Institute for Solar-Terrestrial Physics, German Aerospace
    Center (DLR), Kalkhorstweg 53, D-17235 Neustrelitz, Germany} }

\abstract{We present a general framework on how the polarization of
  radiation due to scattering, dichroic extinction, and birefringence
  of aligned spheroidal dust grains can be implemented and tested in
  3D Monte Carlo radiative transfer (MCRT) codes. We derive a
  methodology for solving the radiative transfer equation governing
  the changes of the Stokes parameters in dust-enshrouded objects. We
  utilize the M\"uller matrix, and the extinction, scattering, linear,
  and circular polarization cross sections of spheroidal grains as
  well as electrons.  An established MCRT code is used and its
  capabilities are extended to include the Stokes formalism. We compute
  changes in the polarization state of the light by scattering,
  dichroic extinction, and birefringence on spheroidal grains. The
  dependency of the optical depth and the albedo on the polarization
  is treated. The implementation of scattering by spheroidal grains
  both for random walk steps as well as for directed scattering
  (peel-off) are described.  The observable polarization of radiation
  of the objects is determined through an angle binning method for
  photon packages leaving the model space as well as through an
  inverse ray-tracing routine for the generation of images.  We
  present paradigmatic examples for which we derive analytical
  solutions of the optical light polarization by spheroidal dust
  particles. These tests are suited for benchmark verification of
  {\name} and other such codes, and allow to quantify the numerical
  precision reached.  We demonstrate that {\name} is in excellent
  agreement to within $\sim 0.1$\% of the Stokes parameters when
  compared to the analytical solutions.}

\date{Received October 25, 2021 / Accepted ???, ???}

\maketitle


\section{Introduction}

Dust processes the radiation that astronomical objects
emit. Especially ultraviolet to optical light is easily scattered or
absorbed. The dust also emits radiation, mostly at infrared to
sub-millimetre wavelengths.  Nearly all astronomical objects are seen
through dust shrouds and their spectral energy distributions (SED) are
altered by this. Dust also polarizes the radiation that is scattered,
extinct or emitted by grains while passing through the
medium. Observations have shown the polarization of radiation due to
dust, for example around active galactic nuclei \citep{Miller1990},
around supernovae \citep{Tran1997}, around single stars
\citep{Forrest1975}, in the galactic interstellar medium
\citep{Serkowski1975}, or other galaxies \citep{Montgomery2014}.

Dust clouds have complex morphologies and varying density profiles,
and they contain non-spherical dust grains of various sizes and
compositions, which are partially aligned. An analytic description of
the signatures of dust on the radiation field is only possible under
the assumption of strong simplifications, which generally do not
hold. A common numerical technique to account for all these different
dependencies of the dust--photon interactions is the Monte Carlo
radiative transfer (MCRT) approach. In this procedure, photon packages
propagate probabilistic through a simulation of the volume under
study. Effects like absorption, scattering, and emission of photons
are explicitly carried out and the effects on the dust temperature are
recorded. A vast body of theoretical work has been developed for MCRT,
a starting point can be the review by \citet{Steinacker2013}.  A large
number of radiative transfer codes based on the Monte Carlo technique
are available, and their scopes, sophistication, and application
domains are extremely different.

The polarization of radiation influences how it interacts
with the dust and the interaction with the dust influences the
polarization.  Several MCRT codes, therefore, started treating
polarization in simplified schemes. A first step is often taken by
calculating the polarization which is due to scattering at spherical
dust grains and electrons. MCRT codes that include such polarization
mechanisms have been presented by \citet{Voshchinnikov1994},
\citet{Bianchi1996}, \citet[][TORUS]{Harries2000},
\citet[][Pinball]{Watson2001}, \citet[][MCFOST]{Pinte2006},
\citet[][MCMAX]{Min2009}, \citet[][HYPERION]{Robitaille2011},
\citet[][STOKES]{Goosmann2014}, and \citet[][RADMC-3D]{Kataoka2015},
to name a few. We have also treated polarization due to scattering on
dust as presented in \citet{Peest2017}, hereafter called
\citetalias{Peest2017}.

Very few MCRT codes implement additional and more sophisticated
processes that can lead to the polarization of radiation by
dust. Among them are codes developed by \citet{Whitney2002} and
\citet{Lucas2003}, which treat scattering, dichroic extinction and
birefringence due to perfectly aligned spheroidal grains. The POLARIS
code \citep{Reissl2016} calculates the polarized emission, scattering,
dichroic extinction and birefringence due to (imperfectly) aligned
oblate grains. The MCRT code by \cite{Bertrang2017} uses in the first
step spherical grains for the dust heating and scattering processes
and then it uses non-spherical grains aligned by radiative torques and
magnetic fields for the dust emission phase. MoCafe
\citep{Lee2008,Seon2018} uses spherical grains including polarization
by scattering and an empirical formula to emulate dichroism based on
optical depth and magnetic field alignment. \citet{Vandenbroucke2021}
implemented polarized emission by partially aligned spheroidal grains
in the SKIRT MCRT code \citep{Camps2020}.

A code that implements or improves an established functionality can be
verified by comparing its results with previous numerical
computations.  Such benchmark tests are available for dust radiative
transfer (RT) models of unpolarized light by \citet{Ivezic1997},
\citet{Pascucci2004}, \citet{Pinte2009}, and \citet{Gordon2017}. They
provide numerical proof for the scattering, extinction, and emission
of radiation due to dust in various environments. For treatments of
dust polarization, such tests are generally not accessible except one,
which covers the case of scattering by spherical dust grains
\citep{Pinte2009}. It compares polarization images of a flared dust
disk around a central star. The dust is spherical and extends
optically thin above and below the optically thick disk. The geometry
of the test provides mostly the accuracy of the polarization after
single scattering events.  As discussed in \citetalias{Peest2017} when
various of the above MCRT codes are applied to more general cases they
show significant differences without the correct results being known a
priori. Flexible and easily reproducible tests are necessary
to benchmark and verify the current and future implementations of
polarization due to non-spherical dust.

In this paper, we present a simple and efficient implementation of the
polarization of radiation due to scattering, dichroic and birefringent
extinction by spheroidal particles. We provide analytical test cases
against which we verify our implementation and estimate the numerical
accuracy of our code, which we call hereafter {\name}. The goal of
this paper is to provide test cases to other teams enabling them to
verify and estimate the numerical precision of their codes. This
should allow groups from many different research areas to explore
polarization including estimated numerical uncertainty.

In Sec.~\ref{p2:sec:Polarization} we present the basic equations
governing dichroic extinction and scattering of radiation by
spheroidal dust grains. We illustrate the functionality of {\name},
and detail how we implement these polarization mechanisms
(Sec.~\ref{p2:sec:Method}). We describe the validation methods in
Sec.~\ref{p2:validateScattering.sec}, which are applied to confirm that the
implementations work as desired. We discuss our results and present
our conclusions in Sec.~\ref{p2:sec:Conclusion}.


\section{Radiative transfer with polarization}
\label{p2:sec:Polarization}

\subsection{Stokes vector and M\"uller matrices}

The RT equation describes the interaction of radiation with
matter. For a medium that absorbs, scatters, and emits radiation, the
basic RT equation reads
\begin{multline}
\label{p2:eq:basicRT}
\frac{\txd I}{\txd s}(\bfr,\bfk)
=
j(\bfr,\bfk)
- n(\bfr)\,\cext(\bfr)\,I(\bfr,\bfk)\\
+ n(\bfr)\,\csca(\bfr) \int \Phi(\bfr,\bfk,\bfk')\,I(\bfr,\bfk')\,\txd\Omega'
\end{multline}
with $I$ the specific intensity of the radiation field\footnote{In our
 notation we do not write obvious wavelength dependencies, e.g., for the
 intensity and the cross sections.}, $n$ the matter density, $j$ the
anisotropic emissivity of radiation, $\cext$ the extinction cross
section, $\csca$ the scattering cross section, and $\Phi$ the phase
function. The different quantities are depending on the position in the
medium $\bfr$, and the direction of the radiation $\bfk$.

This description is incomplete, as it does not consider the
polarization of the radiation. We use the Stokes formalism instead, in
which the radiation field is the 4D Stokes vector $\bfS$,
\begin{equation}
I \quad\Longrightarrow\quad \bfS =
\begin{pmatrix}
I\\
Q\\
U\\
V
\end{pmatrix}.
\end{equation}
The first element describes the intensity of the radiation, the second and third the linear polarization and the fourth the circular polarization \citep{Stokes1852}. There are different conventions in the literature concerning the handedness \citep{Hamaker1996,Peest2017}, we use the convention favored by the IAU \citep{IAU1974}, which for example is not applied by \citet{Ade15}.

Changes to the Stokes vector are described by $4\times4$ M\"uller matrices $\bfM$,
\begin{equation}\label{p2:eq:generalMueller}
\bfS' = \bfM\ \bfS = \begin{pmatrix}
M_{11} & M_{12} & M_{13} & M_{14}\\
M_{21} & M_{22} & M_{23} & M_{24}\\
M_{31} & M_{32} & M_{33} & M_{34}\\
M_{41} & M_{42} & M_{43} & M_{44}
\end{pmatrix}
\begin{pmatrix}
I\\ Q\\ U\\ V
\end{pmatrix}.
\end{equation}
Linear polarization refers to a particular direction, which can be
any direction perpendicular to the propagation direction. In analogy
to observations, we call our choice of the reference direction
{\it{North}}, $\bfd_\N$, which shall not be confused with magnetic
North. In our definition, North is ``up'' when looking against the
propagation direction towards the source. In the plane of the sky East
is counted ``left'' from the North as is common in astronomy.  The Stokes
vector changes depending on the choice of North. When rotating the
North direction by an angle $\beta$, the Stokes vector must be
multiplied with a rotation matrix $\bfR(\beta)$
\begin{equation}
\bfS' =  \bfR(\beta)\,\bfS\ =
\begin{pmatrix}
1 & 0 & 0 & 0 \\
0 & \cos2\beta & \sin2\beta & 0 \\
0 & -\sin2\beta & \cos2\beta & 0 \\
0 & 0 & 0 & 1
\end{pmatrix}
\,\bfS .
\label{p2:eq:rotation}
\end{equation}

The ideal choice of reference direction depends on the phenomenon.

\subsection{Scattering by spheroidal grains}

For radiation that is scattered on spherical grains, the North is usually
chosen in the plane of scattering, defined by the propagation
direction before scattering $\bfk$ and the propagation direction of
the photons after scattering $\bfk'$ \citep[see
  e.g.,][]{Chandrasekhar1960}. This allows for more efficient
approaches when treating scattering events as the 16 entries of the
scattering matrix (see below) can be simplified in the case of
spherical grains to just four independent elements. The method is
widely applied \citep[e.g.,][]{Goosmann2007}.

For spheroidal grains, the M\"uller matrices typically have their
simplest form when the North is in the plane of incidence
\citep{Mishchenko2002}, defined by the propagation direction before
scattering and the grain symmetry axis
(Fig.~\ref{p2:fig:scattering_geometry}).

Scattering events by spheroidal grains are described by the $4\times4$
M\"uller matrix denoted as $\bfZ$ and called the scattering
matrix. For each incoming photon $\bfk$, $\bfZ(\bfk,\bfk')$ provides
the probability that it is scattered towards an outgoing direction
$\bfk'$. The matrix elements depend on the grain shape, size,
porosity, and optical constants, as well as the wavelength, and direction
of incoming and outgoing radiation.  The calculation of $\bfZ$ can be
simplified using symmetry arguments. in the case of spheroidal grains
in $\bfZ$ only seven of the 16 elements are independent
\citep{Bohren1998, Abhyankar1969}. For spheroids, the separation of
variables method has been established \citep{Asano1975,
  Voshchinnikov1993}. For rotationally symmetric particles, the
scattering matrix can be calculated using the so-called T-matrix
method \citep{Mishchenko1996, Mishchenko1991b, Vandenbroucke2020}. For
general non-spherical particles, $\bfZ$ can be calculated by binning the
grains into a grid of dipoles \citep{Purcell1973, Draine1988,
  Draine1994}. Several codes are publicly available or can be
requested from the above-listed authors. The different methods provide
consistent results in computing the cross sections up to size
parameter $x = 2 \pi \/ a/\lambda \sim 10$ and encounter numerical
problems at $x \simgreat 20$ \citep{S19, DH21}.

Because of the rotational symmetry of spheroids, the direction of the
incoming radiation can be described by one angle, the angle of
incidence. It is defined as the angle between the direction of
propagation and the grain symmetry axis.

The scattering part of the RT equation (Eq.~\ref{p2:eq:basicRT})
changes to a tensor equation when polarization is considered
\begin{multline}
\label{p2:eq:scatteringRT}
n(\bfr)\,\csca(\bfr) \int \Phi(\bfr,\bfk,\bfk')\, I(\bfr,\bfk')\, \txd\Omega'
\\ \Longrightarrow\quad
n(\bfr)\int \bfZ(\bfr,\bfk,\bfk')\, \bfS(\bfr,\bfk')\,\txd\Omega'
\end{multline}
Following \citet{Whitney2002} and \citet{Mishchenko2002} we can
calculate the total scattering cross-section
$\tilde{C}_\text{sca}$. It is defined as the integral over all the
scattered intensity $I'$ relative to the intensity $I$
incident onto a spheroidal grain,
\begin{align}
\tilde{C}_\text{sca}
&= \frac{1}{I}\int I'\,\txd\Omega' \nonumber \\
&= \frac{1}{I}\int \left(Z_{11}I + Z_{12}Q + Z_{13}U + Z_{14}V\right)\txd\Omega' \nonumber \\
&= \int Z_{11}\,\txd\Omega' + \frac{Q}{I}\int Z_{12}\,\txd\Omega' \nonumber \\
&= \csca + \cscapol\,\frac{Q}{I}.
\label{p2:eq:scaCrosSec}
\end{align}
The integrals over $Z_{13}$ and $Z_{14}$ are zero, because the grains
are rotationally symmetric and we integrate over the entire unit
sphere \citep[][p.\ 47--51]{VanDeHulst1957}. The ``classical''
scattering cross section for unpolarized radiation is $\csca$, while
$\cscapol$ is the polarization scattering cross-section, which
complicates matters and becomes important for polarized impinging
light. Hence, the probability that a photon is scattered by a spheroid
depends on the polarization status of the photon.

\subsection{Extinction by spheroidal grains}

Extinction by spherical grains is very straightforward and fully
described by a single quantity, the extinction cross-section
$C_{\text{ext}}$. For spheroidal grains, things are complicated, as the
cross-section also depends on the polarization status of the
radiation. The relevant effects are called dichroism and
birefringence. Dichroism means that the extinction differs for
differently polarized radiation. Birefringence means that the travel
speed of the photons through the medium depends on their polarization
status.

The extinction term in the RT equation can be expanded to contain the dichroism and birefringence effects,
\begin{equation}
-n(\bfr)\,\cext(\bfr)\,I(\bfr,\bfk) \quad\Longrightarrow\quad
-n(\bfr)\,\bfK(\bfr,\bfk)\,\bfS(\bfr,\bfk)
\end{equation}
with $\bfK$ the extinction matrix. For spheroids with North in the
plane of incidence, the extinction matrix has a block diagonal shape
\citep{Martin1974, Mishchenko1991, Whitney2002, Lucas2003,
  K08, Voshchinnikov2012},
\begin{equation}
\bfK  =
\begin{pmatrix}
\cext & \cpol & 0 & 0\\
\cpol & \cext & 0 & 0\\
0 & 0 & \cext & \ccpol\\
0 & 0 & -\ccpol & \cext
\end{pmatrix}.
\label{p2:eq:extinctionMatrix}
\end{equation}
The ``classical'' extinction cross section is $\cext$. The dichroism
cross section $\cpol$ describes how much more a radiation wave is
polarized parallel to the North direction is extincted than a
radiation wave polarized parallel to the East direction.  The
birefringence cross-section $\ccpol$ represents how much more a North
polarized wave is slowed down than an East polarized wave. All cross
sections depend on the grain shape, size, porosity, and optical
constants, the wavelength, and the angle of incidence.

\subsection{RT equation}

When polarization due to spheroids is considered, the basic RT
equation~\eqref{p2:eq:basicRT} becomes the following partial
integrodifferential vector equation
\begin{multline}
\label{p2:eq:polarizedRT}
\frac{\txd\bfS}{\txd s} (\bfr,\bfk)
= j(\bfr,\bfk)
- n(\bfr)\,\bfK(\bfr,\bfk)\,\bfS(\bfr ,\bfk)
\\
+ n(\bfr) \int \bfZ(\bfr,\bfk,\bfk')\,\bfS(\bfr,\bfk')\,\txd\Omega'.
\end{multline}
The goal of polarized radiative transfer is to derive the Stokes
vector $\bfS$ at any position $\bfr$ and in any direction $\bfk$ for a
given density $n$, emissivity $j$ and optical properties and alignment
of the dust grains.


\section{Monte Carlo solution of polarized radiative transfer}\label{p2:sec:Method}

\subsection{{\name}}

We implement polarization in the MCRT code developed by
\citet{K08}.  The efficiency of the code was significantly
improved by \citet{Heymann2012} who vectorized it using graphical
processing units (GPUs) and applied for optically thin cells the
treatment by \citet{Lucy1999} and for optically thick cells the method
by \citet{Fleck1984}.  A ray-tracing routine allows the computation of
images of areas of interest at any wavelength. It uses the scattering
and absorption events treated in the simulation and enable the
calculation of SEDs of subdomains of the model.  The code has been
applied to describe the effective extinction curve when photons are
scattering in and out of the observing beam \citep{Kruegel2009}. It
has been used to investigate the structure of disks around T Tauri and
Herbig Ae stars \citep{Siebenmorgen2012}, to create a two-phase AGN
SED library \citep{Siebenmorgen2012}, to study the effects of a clumpy
ISM on the extinction curve \citep{Scicluna2015b}, and to examine the
the appearance of dusty filaments at different viewing angles
\citep{Chira2016}. A time-dependent MCRT version was used to discuss
the impact of a circumstellar dust halo on the photometry of
supernovae~Ia \citep{Kruegel2015}.

So far polarization has not been considered in {\name}. We present a
MCRT dust polarization implementation for spheroidal grains that keeps
the code backwards compatible. This means that the logical order of the
 processing steps remains as before and that all calculations required
for the polarization  are in a module separate from
the main program. The most important change compared to the previous
MCRT code is that we add the Stokes formalism. Each photon package in
the original version of the code was characterized by its frequency,
origin, and propagation direction; in the new version, the East
direction $\bfd_\E$ and the Stokes $Q$, $U$ and $V$ parameters are
stored as well, as motivated in \citetalias{Peest2017}.

In the following subsections, we describe how the effect of scattering
and extinction by spheroidal grains is implemented in {\name}. First
the Stokes vector must be oriented so that North is in the plane of
incidence for the aligned dust grains
(Sec.~\ref{p2:sec:orientStokesVec}).  A photon package propagating
through a cloud of spheroids continuously changes its Stokes vector
(Sec.~\ref{p2:sec:propPhoton}).  The relationship between physical
path length through a cloud of aligned spheroids and the optical depth
experienced by radiation is discussed in Sec.~\ref{p2:sec:tauFromDs}.
In case the radiation interacts with the dust, a part of the intensity
is either scattered or absorbed following its albedo. The albedo of
the dust depends on the polarization of the incoming radiation, which
further complicates the problem (Sec.~\ref{p2:sec:calcAlbedo}). If a
scattering takes place, the propagation direction of the photon after
the scattering is obtained via rejection sampling of the scattering
M\"uller matrix (Sect.~\ref{p2:sec:rejectionSampling}). The
polarization of photons exiting the model space is recorded
(Sec.~\ref{p2:sec:detectPhoton}). Additionally, we implement a routine
to calculate the probability of directed scattering (peel-off,
Sec.~\ref{p2:sec:peel-off}).  Finally, a routine is presented that is
capabable of treating dichroism for inverse ray-tracing, which can be
used for the generation of polarization maps
(Sec.~\ref{p2:sec:raytracer}).


\subsection{Orienting the Stokes vector}
\label{p2:sec:orientStokesVec}

The polarization of the radiation changes as it propagates through the
model space. As described in Sec.~\ref{p2:sec:Polarization}, these
changes are encoded in M\"uller matrices. When dealing with aligned
spheroidal grains, the M\"uller matrices have their simplest form when
the North direction of the radiation is in the plane of incidence,
defined by the propagation direction of the radiation $\bfk$ and the
 symmetry axis of the grains $\bfo$.

The propagation direction $\bfk$, the North direction $\bfd_N$, and
the East directions $\bfd_\E$ are unit vectors and describe a right
handed coordinate system (Fig.~\ref{p2:fig:scattering_geometry}),
\begin{equation}
\bfd_\E = \bfk \times \bfd_\N  \qquad \mathrm{and} \qquad \bfd_\N = \bfd_\E \times \bfk \ .
\end{equation}
During the life cycle of a photon package, the orientation of the
plane of incidence alters frequently: it changes either when the
orientation of the grain’s changes, e.g., when a photon package enters a
 new cell where the grains are oriented in a different way as in the
previous cell or when the propagation direction changes e.g., after a
scattering event or when a new photon gets emitted by the dust. When
either of these situations occurs, we need to rotate the Stokes vector
to ensure that the North direction is in the plane of incidence.

We compute the normal to the plane of incidence $\bfn$ from the
propagation direction $\bfk$ and the grain symmetry axis $\bfo$,
\begin{equation}
\bfn = \frac{\bfk \times \bfo}{|\, \bfk \times \bfo \, |}
\label{ninc}
\end{equation}
Let $\beta$ be the angle that describes the rotation so that the North
direction is in the plane of incidence. Then $\beta$ is also the angle
such that the East direction is perpendicular to the plane of
incidence. Therefore $\beta$ is the angle between $\bfd_\E$ and
$\bfn$. We calculate $\beta$ from the two directions and
the vector algebra relations,
\begin{align}
\cos \beta &= \bfd_\E \cdot \bfn\\
\sin \beta &= \bfk\cdot\left(\bfd_\E \times \bfn \right)
\end{align}
We then rotate the Stokes vector $\bfS$ to the plane of incidence using $\beta$,
\begin{equation}
\bfS' =  \bfR (\beta)\ \bfS
\end{equation}
We would like to stress again that the plane of incidence and the
angle of incidence depend on both $\bfk$ and $\bfo$. The steps above
must be repeated whenever either of them changes, that is
when the photon is scattered or enters a cell with a different grain
orientation or when a new photon is emitted. Note that Eq.~\ref{ninc}
breaks down and cannot be applied when the angle of incidence is
0\degree\ or 180\degree, that is, when $\bfk$ and $\bfo$ are parallel
or anti-parallel. In this case, the photon path is oriented along the
spheroid's symmetry axis, and there is no preferential direction for
the North or East direction. In that case, any arbitrary direction
perpendicular to $\bfk$ can be chosen as $\bfn$.


\subsection{Propagating the photon package}
\label{p2:sec:propPhoton}

The polarization of a photon changes when travelling through a dichroic
or birefringent medium. To mathematically describe this, we consider a
 beam of light propagating through such a medium. Without dust emission
or scattering, the polarized RT equation~\eqref{p2:eq:polarizedRT}
simplifies to
\begin{equation}
\frac{\txd\bfS}{\txd s}(\bfr,\bfk)
= -n(\bfr)\,\bfK(\bfr,\bfk)\,\bfS(\bfr,\bfk).
\end{equation}
As we assume that the density, grain orientation, and the optical
properties within a single dust cell are constant, this linear
differential matrix equation can be solved analytically within each
individual cell. Using Eq.~\eqref{p2:eq:extinctionMatrix} we obtain
two coupled systems and their solution is \citep{Lucas2003,
  Whitney2002}
\begin{equation}
\label{p2:eq:dichroicExtinction}
\bfS(s) = \e{-\cext n  s}
\begin{pmatrix}
I_0\cosh( \cpol n  s ) - Q_0\sinh( \cpol n  s )\\
Q_0\cosh( \cpol n  s ) - I_0\sinh( \cpol n  s )\\
U_0\cos( \ccpol n  s ) - V_0\sin( \ccpol n  s )\\
V_0\cos( \ccpol n  s ) + U_0\sin( \ccpol n  s )
\end{pmatrix}.
\end{equation}
The equation describes the change of the Stokes vector when the photon
travels through a single cell in a dichroic or birefringent
medium. The change of the Stokes vector $\bfS(s)$ needs to be
considered in MCRT computation when the photon packets propagate by
$s$ in such a medium. We note that, whenever the photon package enters
a new cell where the grain orientation is different, we need to apply
a re-orientation of the Stokes vector
(Sec.~\ref{p2:sec:orientStokesVec}) before we can continue the
propagation.


\subsection{Optical depth and path length}
\label{p2:sec:tauFromDs}

\begin{figure}
\centering
\includegraphics[width=9cm,clip=true,trim=4cm 8cm 3.5cm 8.8cm]{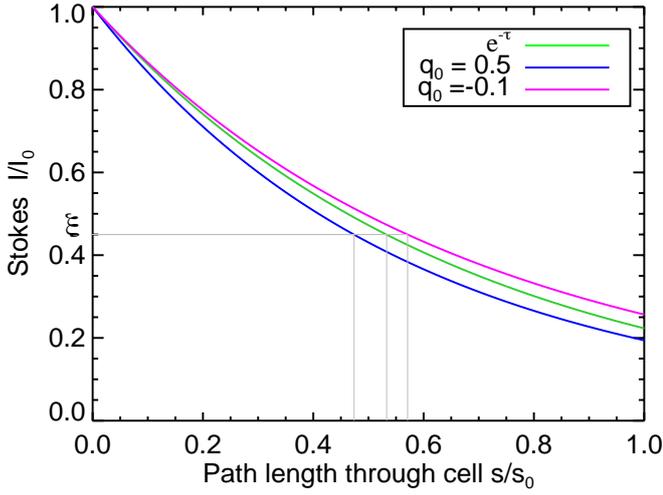}

\caption{Change of Stokes intensity along a flight path of a photon
  along a cell.  The decay of the intensity has been shown for
  unpolarized light (green) and polarized light
  (Eq.~\ref{p2:eq:dichroicExtinction}) with $q_o = Q_o/I_o = 0.5$
  (blue) and $q_o = -0.1$ (magenta). The interaction points of light
  with dust for random incidence $\xi = 0.45$ are indicated by the
  grey lines. }
  \label{p2:fig:dichroism_extinctionCurves} \end{figure}

One necessary exercise in MC treatments is computing the optical depth
along the flight path of the photon package through the cell for
determining whether and where it will interact with the dust. In
{\name} the grid cells are cubes with constant density $n$ and grains
have a fixed alignment. The flight path of the photon packets go along
a straight line from the entry point into a cube up to either the
interaction point or in case of no interaction up to the exit point of
that cube.

Photon packets of unpolarised light in a non-dichroic medium travel
the distance from the entry point of a cube to the interaction point
$\Delta s$ corresponding to an optical depth $\tau(\Delta s$). This
optical depth can be computed from a random exponential distribution
such that the interaction of the photon package with dust in a cell
follows a uniform distributed random number $\xi$ \citep{K08,
  Witt1977, Lucy1999, Steinacker2013},

\begin{equation}\label{snopol.eq}
  \Delta s = \frac{- \ln (\xi)}{n \cext} \ .
\end{equation}

The optical depth along $\Delta s$ is

\begin{equation} \label{p2:eq:classicTau}
  \tau(\Delta s) = \cext \, n \, \Delta s \ .
\end{equation}

In comparison to the unpolarized light, the photon packets of
polarised light in a dichroic medium travel the distance from the
entry point of a cube to the interaction point $\Delta s^{\rm pol}$
and corresponding optical depth $\tau^{\rm pol}(\Delta s^{\rm
  pol}$). We add suffix $\rm{pol}$ for quantities that depend on
polarization.  Following \cite{Baes19}, a direct connection between
the physical path length and the optical depth as given for
unpolarised light is lost in media with dichroism and
birefringence. Considering polarization in a dichroic medium the
optical depth cannot be calculated in closed form as in
Eq.~\eqref{p2:eq:classicTau}. Diminishing of the polarized radiation
needs to consider the change of the Stokes vector {\bf{${S}$}} when
the photon packet travels through that cell and follows
Eq.~\eqref{p2:eq:dichroicExtinction}. In that solution, there is
besides the extinction, which is given by an exponential decay, these
additional terms of the Stokes components noted between the
brackets. We visualize in Fig.~\ref{p2:fig:dichroism_extinctionCurves}
that dichroism indeed complicates the relation between path length and
optical depth. We choose two examples of the exact solution for
polarized light of Eq.~\ref{p2:eq:dichroicExtinction} adopting $q_o =
Q_o/I_o = 0.5$ and $q_o = -0.1$ and show the solution for unpolarized
light given by $\exp \left( -\tau\right)$. For better visualization,
an extremely dichroic cell has been adopted with $n = 0.0015$, $\cext
= 1000$, and $\cpol = 300$\,(a.u.), hence the extinction optical depth
$\tau(1) = 1.5$.  Note the vast variations of the interaction points
${\Delta s}$ and ${\Delta s^{\rm {pol}}}$ for the same optical depth
either $\tau$ or $\tau^{\rm {pol}}$ at same random number $\xi$. The
interaction points are at $\Delta s^{\rm {pol}} = 0.47$ for $q_o =
0.5$, $\Delta s^{\rm{pol}}=0.57$ for $q_o = -0.1$ and $\Delta s =
0.54$ for unpolarized light
(Fig.~\ref{p2:fig:dichroism_extinctionCurves}).

To stress the fact that we deal with dichroic extinction, we use the
term polarization optical depth $\tau^{\rm {pol}}$.  We use the
definition of an optical depth and insert
Eq.~\eqref{p2:eq:dichroicExtinction}

\begin{align} \label{p2:eq:deltaTauDC}
  \tau^{\rm {pol}}(s) &= -\ln\left(\frac{I(s)}{I(0)}\right)\\
                & = \cext \, n  \, s \ - \ln \left( \cosh( \cpol \, n \, s)- \sinh( \cpol \, n \, s)\frac{Q_0}{I_0} \right)
   \label{p2:eq:deltaTauDC2}
\end{align}

The right-hand side of Eq.~\ref{p2:eq:deltaTauDC2} can, except for
special cases, not be simplified. For non-spherical dust particles,
there is another complication that the orientation of the Stokes
vector to the grain orientation, hence the dust cross sections need to
be considered. In {\name} the grain alignment in a cell is fixed,
which simplifies computing the angle of incidence $\alpha$, which is
the angle between the Stokes vector of the incoming photon and the
grain orientation (Fig.~\ref{p2:fig:scattering_geometry}). The Stokes
vector of the incoming photon needs to be rotated to the plane of
incidence and $\cext (\alpha)$ and $\cpol (\alpha)$ are computed for
that $\alpha$. For clarity, the angle and frequency dependence has
been dropped in Eq.~\ref{p2:eq:deltaTauDC2}.

Whenever taking dichroism into account the polarization state changes
along the flight path of the photon (Sect.\ref{p2:sec:propPhoton}) and
an analytical solution for $\Delta s^{\rm {pol}}$ does not
exist. \citet{Baes19} showed that the solution can be expressed by a
Taylor expansion and that the extinction optical depth $\tau$, this is
the optical depth ignoring polarization approximates the polarization
optical depth $\tau^{\rm {pol}}$ to first order, hence

\begin{equation}\label{spolin.eq}
\frac {\Delta s} {\Delta s^{\rm {pol}}} \approx
\frac{\tau (\Delta s)} {\tau^{\rm {pol}}(\Delta s)}\ .
\end{equation}

\noindent

The density $n$ of the cube, the angle of incidence $\alpha$, and
$\cext(\alpha)$ are known. As indicated by the horizontal line in
Fig.~\ref{p2:fig:dichroism_extinctionCurves}, the same optical depth
for unpolarised and polarised light are chosen by the same random
number, {${-\ln(\xi) = \tau (\Delta s) = \tau^{\rm{pol}}(\Delta s^{\rm
      {pol}})}$}, which leads to different path lengths $\Delta s \neq
\Delta s^{\rm {pol}}$. The distance ${\Delta s}$ is derived inserting
the quantities $n$, $\cext$, and {$ \tau (\Delta s)=-\ln(\xi)$} into
Eq.~\ref{snopol.eq}. This allows computing ${\tau^{\rm {pol}}(\Delta
  s)} $ using Eq.~\ref{p2:eq:deltaTauDC2}. Hence, we derive the path
length of the polarised light

\begin{equation}\label{spol.eq}
\Delta s^{\rm {pol}} \approx \Delta s \ \frac{\tau^{\rm {pol}}(\Delta s)}{- \ln(\xi)} \ .
\end{equation}

In the magnified examples of
Fig.~\ref{p2:fig:dichroism_extinctionCurves}, the approximation of
$\Delta s^{\rm {pol}}$ (Eq.~\ref{spol.eq}) is within $0.3$\,\% of the
correct solution. For typical ISM dust $\cpol$ is about a factor 100
smaller \citep{S22} and the uncertainty in $\Delta s^{\rm {pol}}$ is
below $10^{-4}$.  Whenever possible, the grid in {\name} is set up so
that the total extinction optical depth of a cube, or when needed
sub-cube, is about $\tau \la 1$. Radiation penetrating through a
highly optically thick medium is challenging for MCRT codes, e.g.,
\citet{Min2009, Siebenmorgen2012, Gordon2017, CampsBaes15}. MC
applications that consider polarization treatments at path length
$\tau^{\rm {pol}} \gg 1$ can progressively solve the radiative
transfer equation along the ray in each cube as outlined by
\citet{Baes19}. This goes in hand with extraordinary boost in
computing time and is not considered.


\subsection{Dust albedo of polarized light}
\label{p2:sec:calcAlbedo}

When the radiation interacts with a dust grain, it is either scattered
or absorbed. The ratio of scattered over extincted intensity defines
the albedo $\Lambda$. We rotate the Stokes vector so that its
component $U=0$, apply
Eqs.~\eqref{p2:eq:scaCrosSec}~-~\eqref{p2:eq:extinctionMatrix}, and
derive the albedo of a spheroid including polarization,
\begin{equation}
\label{p2:eq:albedoDichroism}
\Lambda = \frac{\tilde{C}_\text{sca}}{\tilde{C}_\text{ext}}=\frac{\csca + \cscapol\cdot Q/I}{\cext + \cpol\cdot Q/I}
\end{equation}

This equation can be seen as a more general form of the albedo. For
unpolarized radiation or if the polarization scattering cross-section
and the dichroism cross-section are zero,
Eq.~\eqref{p2:eq:albedoDichroism} reduces to its common form.  We also
note that Eq.~\eqref{p2:eq:albedoDichroism} is independent of
$\ccpol$. This is consistent with the physical meaning of $\ccpol$: It
describes a differential phase shift, a delay, between the North and
East polarized parts of the radiation. Such a delay does not reduce
the intensity of the radiation.

\subsection{Sampling the scattering M\"uller matrix}
\label{p2:sec:rejectionSampling}

\begin{figure}
\centering
\includegraphics[width=\columnwidth, trim={4cm 1.2cm 1.4cm 2.2cm},clip]{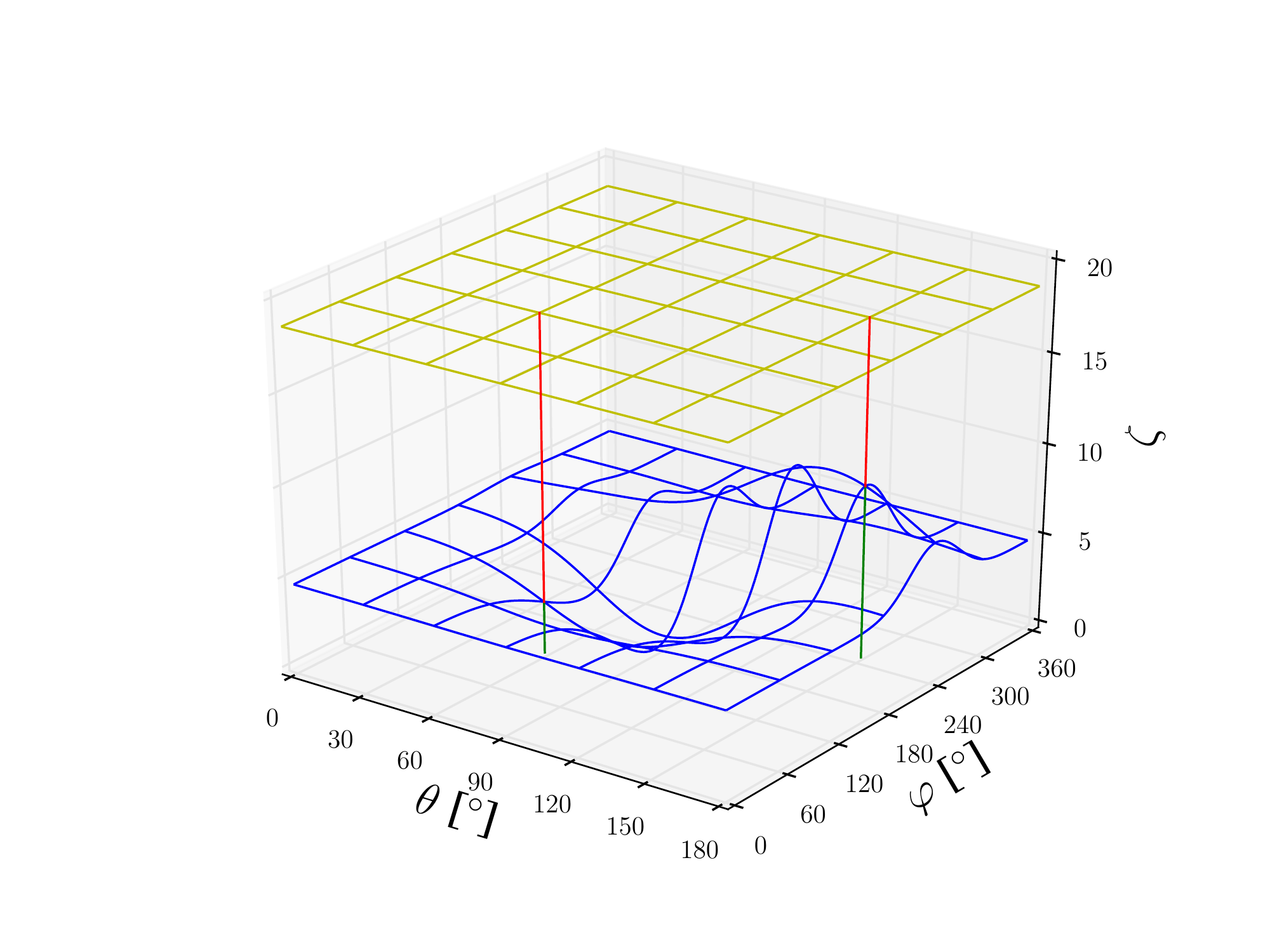}
\caption{Visualization of the rejection sampling method. An arbitrary
  probability density function $P(\theta, \varphi)$ is shown in blue
  and the ceiling value $v_{ceil}$ is in yellow. An angle pair
  $(\theta,\varphi)$ is accepted if a third
  $(0,v_{ceil})$ is smaller than $P(\theta,\varphi)$. For two angle
  pairs, the segments in green and red visualize the $\zeta$ leading to
  acceptance and rejection, respectively. The case where the
  green line is large has a higher chance to be selected than the
  other case where the green line is small. For that selected case of
  the large green line, $P(\theta,\varphi)$ becomes also much larger
  than in the other case as it should be.}
\label{p2:fig:dichroism_rejectionSampling}
\end{figure}

In the case of non-polarized MCRT, simulating a scattering event is
relatively straightforward. It essentially comes down to generating a
random scattering angle $\theta$ from the scattering phase function
$\Phi(\theta)$. Based on the original propagation direction $\bfk$ of
the photon package, this scattering angle can be converted to a new
propagation direction $\bfk'$, and this describes the
scattering event. In polarized MCRT, computing a scattering event is
more complicated: not only is the random generation of a new
propagation direction more complex, but we also must update the
polarization status and the reference direction of the photon package.

Rather than a single scattering angle, we must determine a new pair
of angles $(\theta,\varphi)$. In this pair, $\theta$ is still the
scattering angle, that is the angle between the original propagation
direction $\bfk$ and the new propagation direction $\bfk'$. The angle
$\varphi$ is the azimuth of $\bfk'$ in a coordinate system with $\bfk$
as polar direction and the North direction as the reference direction
with azimuth zero.

The appropriate probability distribution from which the random couple $(\theta, \varphi)$ needs to generated is
\begin{equation}
p(\theta,\varphi) = \frac{I'(\theta,\varphi) \sin\theta}{\int I'(\theta',\varphi') \sin\theta'\,\txd\theta'\,\txd\varphi'}
\label{ppdf}
\end{equation}
where
\begin{multline}
I'(\theta,\varphi) = Z_{11}(\theta,\varphi)\,I + Z_{12}(\theta,\varphi)\,Q
+ Z_{13}(\theta,\varphi)\,U \\
+ Z_{14}(\theta,\varphi)\,V.
\end{multline}
This formula shows two characteristics. Firstly, the
 probability density function is a true bi-variate distribution that
depends explicitly on $\theta$ and $\varphi$ in a nontrivial
way. Secondly, it does not only depend on the characteristics of dust
grains and the grain alignment, but also on the polarization state of
the photon package. To generate a random couple $(\theta,\varphi)$
from the bi-variate probability distribution function (\ref{ppdf}) we
using the rejection sampling method \citep{vonNeumann1951}. This
 method is comparatively simple and is readily applicable to
multivariate probability density functions \citep{Devroye2013,
  Baes2015}. A scattering angle pair ($\theta$, $\varphi$) is generated and
accept if another third random number $\xi_3$ is lower than the
probability $P(\theta, \varphi)$ of scattering in this particular
direction (Fig.~\ref{p2:fig:dichroism_rejectionSampling}). A ceiling
value $v_{ceil}$ gives the maximum probability for any of the angles
and is used to scale the third random number.  The actual
implementation begins with calculating two scattering angles based on
uniform deviates $\xi_i$,

\begin{align}
\varphi &= \xi_1 \cdot 2\pi\\
\label{phisca.eq}
\theta  &= \xi_2 \cdot \pi
\end{align}

It is important to note here, that we sample uniform in $\theta$,
which means that we sample more angles per surface area in the forward
and backward directions. This is motivated by the fact that dust
grains scatter preferably forward or backwards. The sampling density
must be taken into account when we calculate the probability of
scattering in the direction $\theta$, $\varphi$. We calculate a
ceiling value $v_{ceil}$, which is representative for the highest
probability of scattering towards a direction $\varphi$,
$\theta$. Following Eq.~\eqref{p2:eq:generalMueller} with the
scattering matrix $\bfZ$ the outgoing intensity $I'$ for an
 incoming photon with the Stokes vector~$\bfS$ is
\begin{align}
v_{ceil} &= \max\left[ I'(\theta,\varphi)\sin\theta\right]\\
    &=\max\left[\left(Z_{11}I+Z_{12}Q+Z_{13}U+Z_{14}V\right)\sin\theta\right]
\end{align}
with $Z_{ij}(\lambda, \alpha, \varphi, \theta)$ the elements of the
scattering matrix.  The factor $\sin{\theta}$ compensates for
oversampling the forward and backward regions and reducing the ceiling
value $v_\text{ceil}$.  The ceiling value needs to be recomputed for
each scattering event, when the Stokes parameters $Q$, $U$, and $V$,
the angle of incidence $\alpha$, or the wavelength $\lambda$ change.
The decision of whether the angle pair is accepted is based on the third
random number,
\begin{equation}
\zeta   = \xi_3 \cdot v_{ceil}
\end{equation}
with $0 \leq \xi_3 \leq 1$. The angles $\varphi$ and $\theta$ are accepted, if
\begin{equation}
\zeta \leq I'(\theta,\varphi)\sin\theta
\end{equation}
A low ceiling value $v_{ceil}$, will increase the probability of
accepting an angle pair. Our method of changing the sampling density
is a simple approach to reduce the average number of draws until a
pair is accepted. There can be more sophisticated methods, using the
scattering behavior of the dust mixture used in the simulations.

After generating a random angle pair $(\theta,\varphi)$, we need to
update the characteristics of the photon package. The obvious
characteristic to update is the propagation direction. The new
propagation direction after scattering is calculated following
Eq.~(30) of \citetalias{Peest2017},
\begin{equation}
\bfk'
=
\bfk \cos\theta + (\bfd_\E \times \bfk \cos\varphi + \bfd_\E\sin \varphi)\sin \theta .
\end{equation}
The polarization state of the photon package also needs to be updated
since the scattering event does affect the state. The new Stokes
vector becomes
\begin{equation}
\bfS' = \left(\frac{1}{Z_{11}+Z_{12}Q+Z_{13}U+Z_{14}V}\right) \bfZ\,\bfS,
\end{equation}
where the factor between the brackets guarantees that the total
 intensity of the photon package is conserved during the scattering event.
Finally, also the East direction is updated. The convention we use is
that the new North direction is in the plane of departure after
scattering. This plane is given by the direction after scattering
$\bfk'$ and the symmetry axis of the grain $\bfo$, similar to the
plane of incidence. Therefore, the East direction after scattering is
\begin{equation}
\bfd_{\E}'=\frac{\bfk'\times \bfo}{|\bfk'\times \bfo|}.
\end{equation}
If the outgoing direction and the symmetry axis are parallel or
anti-parallel, the photon package continues its path through the model
space without the need to rotate the Stokes vector.


\subsection{Detection of escaped photons}
\label{p2:sec:detectPhoton}

Eventually, photon packages will leave the model space. We record their
exiting directions $\bfk$ and wavelengths. Photons that exit with the
same wavelength and with similar angles to the simulation $z$-axis are
binned together. For axisymmetric geometries, this method allows for
the quick calculation of the spectral energy distribution of the model
under different viewing angles. The Stokes vector of the photon
packages is rotated upon leaving, such that North is in the plane
given by the escape direction and the $z$-axis of the model
space. This permits binning photon packages by adding their Stokes
vectors and intensities.


\subsection{Directed scattering (peel-off)}
\label{p2:sec:peel-off}

MCRT codes commonly allow viewing the model space from (arbitrary)
directions $\bfk_\text{obs}$. This simulates an observation by a
distant observer. The chance of a photon scattering directly towards
the observer is low. We, therefore, employ the peel-off method
\citep{Yusef-Zadeh1984} in which the probability of scattering towards
the observer is calculated explicitly. During the model run, we store
for all scattering events the position, direction $\bfk$ and Stokes
vector $\bfS$ of the photon packages before scattering.  After the
simulation finishes we use the scattering matrix $\bfZ$ to calculate
the Stokes vector $\bfS_\text{obs}$ of the radiation that would have
scattered towards $\bfk_\text{obs}$,
\begin{equation}
\label{p2:eq:peel-off}
\bfS_\text{obs}
= \csca^{-1}\,\bfZ(\varphi_\text{obs}, \theta_\text{obs}) \bfS
\end{equation}
where $\csca^{-1}$ is used as a normalization factor (see below), and
$\varphi_\text{obs}$ and $\theta_\text{obs}$ the angles by which the
photon is scattered towards the observer. The scattering angle
$\theta_\text{obs}$ is calculated from the direction before and after
scattering,
\begin{equation}
\cos\theta_\text{obs} = \bfk \cdot \bfk_\text{obs}
\end{equation}
and the azimuth $\varphi_\text{obs}$ from the East direction
$\bfd_\E$ and the normal to the scattering plane 
calculated from $\bfk$ and $\bfk_\text{obs}$,
\begin{gather}
\cos\varphi_\text{obs} = \bfd_\E \cdot \left( \frac{\bfk \times \bfk_\text{obs}}{|\bfk \times \bfk_\text{obs}|} \right)\\
\sin\varphi_\text{obs} = \bfk \cdot \left( \bfd_\E \times \left( \frac{\bfk \times \bfk_\text{obs}}{|\bfk \times \bfk_\text{obs}|} \right)\right)
\end{gather}
The normalization $\csca^{-1}$ of the scattering matrix is essential
for calculating the peel-off probability using
Eq.~\eqref{p2:eq:peel-off}. The scattering matrix is stored internally
as a multi-dimensional array, with $N_\theta$ and $N_\varphi$ elements
along the respective angles $(\theta, \varphi)$. Following
Eq.~\eqref{p2:eq:scaCrosSec}, the sum of the $Z_{11}$ elements over
the unity sphere is $N_\theta \ N_\varphi \ \csca$. The scattering
probability has already been considered during the MC run and so one
divides by $\csca$.

The intensity of the radiation that is scattered towards the observer
is then the first component of the Stokes vector resulting from
Eq.~\eqref{p2:eq:peel-off}. The intensity of the radiation that would
reach the observer has to be reduced to account for (dichroic)
extinction between the point of scattering and the observer. For
applying this properly we developed an inverse ray-tracing routine
discussed in the following section.


\subsection{Inverse ray-tracing}
\label{p2:sec:raytracer}

An inverse ray tracer is developed for calculating spatially resolved
maps of the model space. For this, it is necessary to know the optical
depth of a scattering event towards the observer. We compute a
Stokes map by sending rays in the direction $-\bfk_\text{obs}$ from
the observer to an area of the model. Along the path of a ray, it
encounters cells that are numbered by $i=1, ..., m$. We calculate for
each cell $i$ the amount of radiation that would be scattered and
emitted towards $\bfk_\text{obs}$ and attenuate this according to the
optical depth $\tau^\text{out}$ from the present position to the entry
point of the ray in the model space. The map is created by varying the
position at which the ray enters the model and under the assumption
that the complete model space is sampled \citep{Heymann2012}.

Without considering polarization, the intensity of the radiation
leaving the simulation is determined by the optical depth to the edge
of the model space, $\tau^\text{out}$. By stepping through the system
along $-\bfk_\text{obs}$ the optical depth increases. The ray crosses
the cells $1,...,m$ and in each cell $i$, the optical depth to the
edge increases by the product of the extinction cross-section of the
dust $C_{\text{ext},i}$, its density $n_i$ and the path length
$(\Delta s)_i$,
\begin{equation}
\tau^\text{out}_i = \tau^\text{out}_{i-1} + C_{\text{ext},i}\,n_i\,(\Delta s)_i.
\end{equation}
The radiation that is scattered or emitted in cell $i$ towards the
 observer with an intensity $I_i$, will exit the model and reach the
observer with the reduced intensity $I_i\,\e{-\tau^\text{out}_i}$.

When we take dichroism into account, the optical depth through cell
$i$ depends on the polarization of the radiation. In addition, the
polarization of the radiation will change along
$\bfk_\text{obs}$. Both effects are described by
Eq.~\eqref{p2:eq:dichroicExtinction}.  One also has to consider the
orientation of the grains, as they can be different for each
cell. Consider the change of the Stokes vector $\bfS$ of the photon
package on its way from cell $i$ along direction $\bfk_\text{obs}$ out
of the model space. It is given by an initial rotation into the frame
of the cell, $\bfR$, and then an alternating application of a dichroic
extinction step towards the edge of that cell, followed by a rotation
into the plane of incidence of the next cell, and so forth until it
eventually leaves the model space. This can be written as a single
equation by combining Eq.~\eqref{p2:eq:dichroicExtinction} and
Eq.~\eqref{p2:eq:rotation},
\begin{equation}\label{p2:eq:DCraytrace}
\bfS^\text{out}_i
=
\bfR_{\text{obs}}\,
\bfR_1\,\e{-\tau_{\text{ext},1}}\,\bfE_1\,\cdots\,\bfR_i\, \e{-\tau_{\text{ext},i}}\,\bfE_i\,\bfR\,\bfS .
\end{equation}

For each cell $j$ with $0\leq j \leq i$, the rotation into the cell
$j-1$ is $\bfR_j$, and $\bfR_{\text{obs}}$ is the rotation of North
into the reference frame of the observer. The optical depth inside $j$
is $\tau_{\text{ext},j} = C_{\text{ext},j}\,n_j\, (\Delta s)_j$ and
the dichroism and birefringence matrix is $\bfE_j$,
\begin{equation}
\bfE_j =
\begin{pmatrix}
\cosh \tau_{\text{pol},j} & -\sinh \tau_{\text{pol},j} &0&0\\
-\sinh \tau_{\text{pol},j} & \cosh \tau_{\text{pol},j} &0&0\\
0&0& \cos \tau_{\text{cpol},j} & -\sin \tau_{\text{cpol},j}\\
0&0& \sin \tau_{\text{cpol},j} & \cos \tau_{\text{cpol},j}
\end{pmatrix}
\end{equation}
where for cell $j$ the dichroism and birefringent optical depths are given by
\begin{gather}
\tau_{\text{pol},j} = C_{\text{pol},j}\,n_j\, (\Delta s)_j, \\
\tau_{\text{cpol},j} = C_{\text{cpol},j}\,n_j\,(\Delta s)_j.
\end{gather}
The ray-tracing equation, Eq.~\eqref{p2:eq:DCraytrace}, assumes that
the photon package is emitted or scattered at the edge between the
cell $i$ and cell $i+1$. This is correct for optically thin cells
($\tau<0.1$). In the case of optically thicker cells, the emission is
distributed along the path in cell $i$ and needs to be integrated
within cell $i$.

The advantage of Eq.~\eqref{p2:eq:DCraytrace} is that it can be
evaluated while stepping along the pencil beam. The product of the
matrices of the previous cells is sufficient to calculate the Stokes
vector of the next cell. For cell $i$, the matrix of the previous cell
$i-1$ is multiplied by $\bfR_i\, \e{-\tau_{\text{ext},i}}\,\bfE_i$.
One can store the compound matrix that is updated when entering
the next cell. The compounded exponential factor of
Eq.~\eqref{p2:eq:DCraytrace} can be stored separately to keep the
entries of the matrix closer to unity and to prevent numerical
instabilities.


\begin{figure}
\centering\includegraphics[width = \columnwidth, trim={0.3cm 2.0cm
    3.8cm 6.2cm},clip]{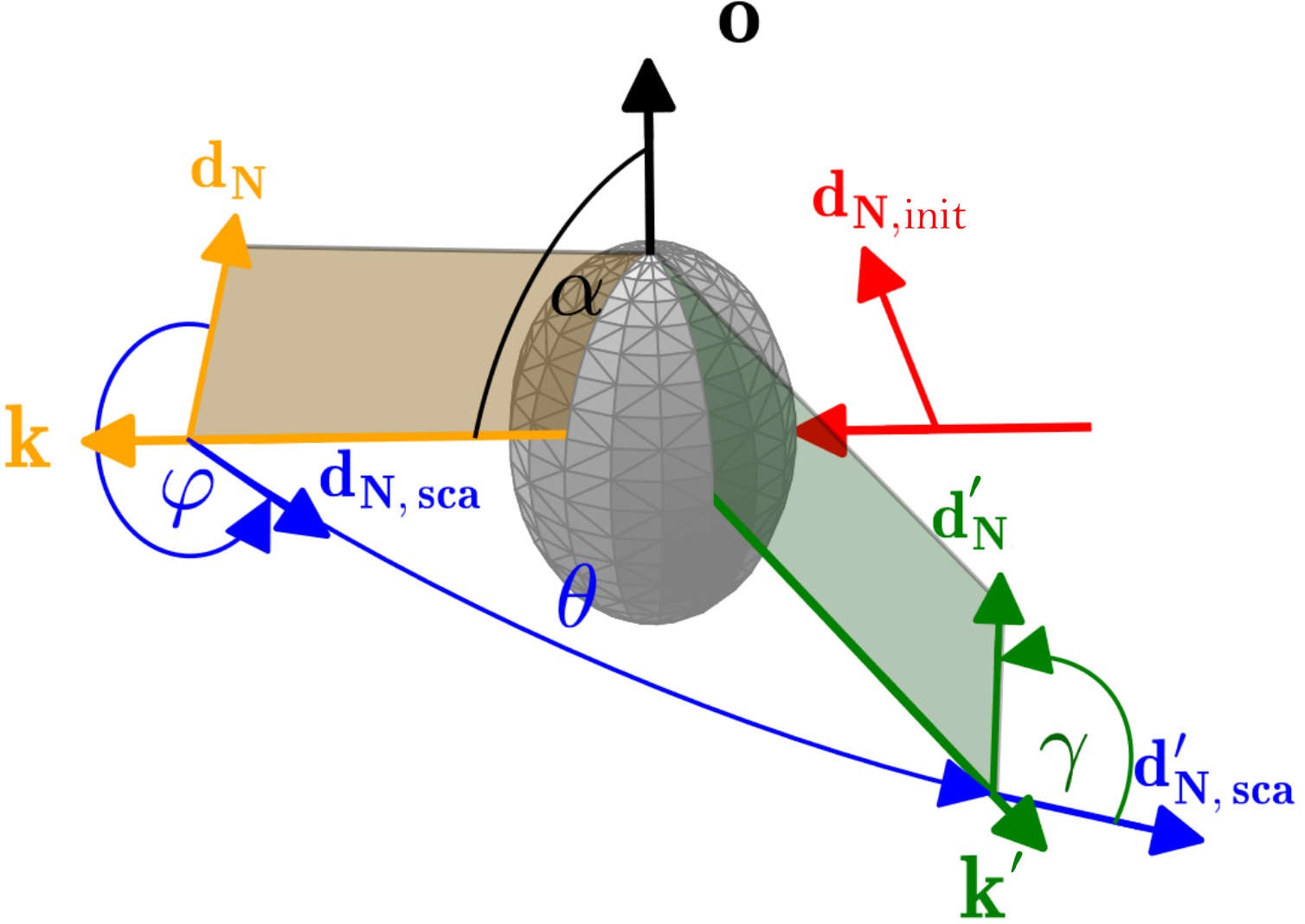}
\caption{Geometry of a scattering event as used for the analytical
  validation in Sec.~\ref{p2:validateScattering.sec}. A photon arrives from
  the right side at the grain with some initial North direction
  ${\bf{d}}_{N,init}$ (red). Grain symmetry axis $\bfo$ and angle of
  incidence $\alpha$ are shown in black. Photon package initial
  direction $\bfk$, the plane of incidence and the North direction in the
  frame of the grain ${\bf{d}}_{N}$ are shown in orange. Scattering
  angles $\theta$ and $\varphi$ and North directions
  ${\bf{d}}_{N,sca}$, ${\bf{d}}_{N,sca}'$ during the scattering
  process are marked in blue. The direction of the scattered photon
  $\bfk'$, exit angle $\gamma$, plane of departure, and outgoing North
  direction ${\bf{d}}_{N}'$ is shown in green.}
\label{p2:fig:scattering_geometry}
\end{figure}

\section{Validation \label{p2:validateScattering.sec}}

MCRT codes
need to be carefully validated. For doing this, one can often use benchmark
 results from existing codes for comparison. However, as scattering,
dichroism, and birefringence due to spheroidal dust grains are
uncommon capabilities for MCRT codes, there are no such benchmarks to
reproduce.  Ideally, the different functionalities are tested
individually. Such a procedure simplifies the identification of
potential shortcomings or even mistakes and it enables those codes with
different sets of functionalities may reuse the same tests.

As advocated by \citetalias{Peest2017}, it is particularly
advantageous to compare the results of our implementations to analytical
solutions. Analytical solutions are easy to reproduce and can be used
to estimate the errors of the numerical treatment. Analytical
solutions are also of interest to other teams aiming to verify their
MCRT codes. In this Section, we develop analytical test cases to verify
our numerical implementation of scattering, dichroism, and
birefringence due to spheroidal dust. The MCRT treatment of scattering
by spheroidal dust (Sec.~\ref{p2:validateScattering.sec}) is confirmed
using renewed versions of analytical test cases by
\citetalias{Peest2017} developed for spherical grains. Additional
analytical test cases for estimating the numerical accuracy of MCRT
codes that are treating dichroism and birefringence mechanisms are
presented in Sec.~\ref{p2:validateDichroBire.sec}.

\begin{figure}
\centering\includegraphics[width = 6.5cm, trim={0.0cm 0.0cm 0.0cm 0.0cm},clip]{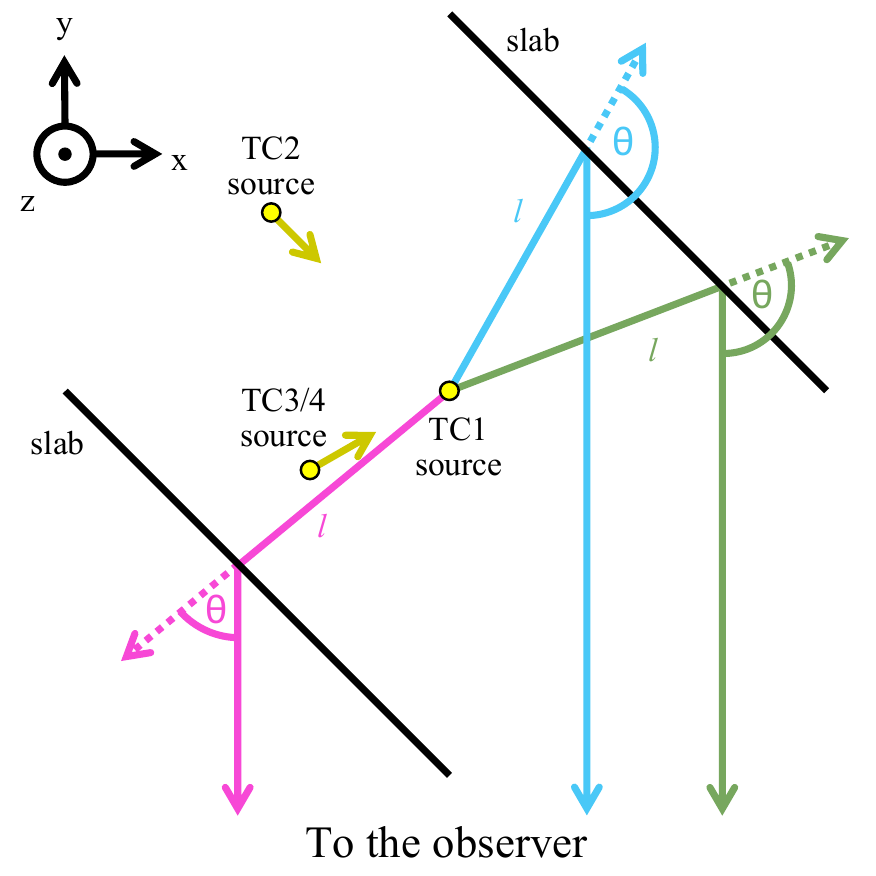}
\caption{Geometry of the analytical test cases. The z-axis is oriented
  towards the reader and the observer towards $-y$. {\it TC~1:} A
  central point source illuminates’ thin slabs of electrons that are
  slanted by 45$^o$ towards the observer. For the other test cases, the
  central point source is replaced by a small cloud of electrons. {\it
    TC~2:} The cloud is illuminated by a collimated beam, which is
  located in the $ xy$ plane. {TC~3 - TC~4:} The illuminating
  source is located below the $xy$-plane.  }
\label{FigTC}
\end{figure}

In \citetalias{Peest2017} we considered radiation scattering on
spherical grains and developed several test cases to validate the
numerical procedure for solving this radiative transfer problem of
polarized light. The analytic solutions can only be expressed exactly
because single scattering in a simplified geometry is considered
(Fig.~\ref{FigTC}) and electrons with their uncomplicated M\"uller
matrix is applied. Four test cases (TC) are distinguished with
analytical solutions given by \citetalias{Peest2017} (Eq.~44-63).
Circular polarization is not considered in TC~1 - TC~3 ($V = 0$) but is
treated in TC~4.

{\it TC~1:} In the first scenario a central point source emits
unpolarized radiation. For testing the peel-off scattering procedure
we select two slabs of electrons. The photons scatter once at these
optically and physically thin electron slabs, which lie in the $ xy$
plane. A distant observer records the intensity, polarization degree
and polarization angle of the radiation scattered off the slabs. TC~1
verifies the polarization by single scattering and the peel-off
mechanism.

In the next test scenarios ({TC~2 - TC~4}), the central light source is
replaced by a tiny cloud of electrons. They are illuminated by a
collimated beam. The analytical solutions of \citetalias{Peest2017}
were derived for photons that encounter a first scattering event at
the center and a second scattering event at the slabs
(Fig.~\ref{FigTC}).

{\it TC~2:} In the second test case the direction of the collimated
beam is in the same plane as the direction of the slabs. The second
scattering is from the electron slab to the observer. In this
configuration, scattering, the peel-off mechanism, as well as a random
walk step of the photons are tested.

{\it TC~3:} In the third test case, the initial beam direction is at
an angle to the plane of the slabs. The plane of scattering of the
initial scattering is rotated to the plane of scattering
 off the slabs. The rotation leads to a variable polarization angle at
the observer.

{\it TC~4:} In the final test case, the scattering properties of the
particles are changed. The M\"uller matrix of the electrons are
hypothetically changed so that radiation scattering on them can become
circular polarized, hence $M_{34} \ne 0$ and $M_{43} \ne 0$
(Eq.~\ref{p2:eq:generalMueller}), which are for electrons zero
otherwise.

We expand the test scenarios of \citetalias{Peest2017}, to cover the
case of scattering at spheroidal dust particles. For spheres and
sphere-like particles, the scattering matrix is most simple when North
is in the scattering plane before and after the scattering event. In
that case, the scattering matrix $\bfZ $ reduces to a block diagonal
shape, which depends for electrons only on the scattering angle
$\theta$ and is independent of $\varphi$, hence

\begin{equation}\label{p2:eq:scatMatSphere}
\bfZ (\theta) =
\begin{pmatrix}
a(\theta)&b(\theta)&0&0\\
b(\theta)&a(\theta)&0&0\\
0&0&c(\theta)&d(\theta)\\
0&0&-d(\theta)&c(\theta)
\end{pmatrix}
\end{equation}

The geometry and our notation of the scattering process on spheroidal
grains is shown in Fig.~\ref{p2:fig:scattering_geometry} and
summarized in Table~\ref{sec.notation}.  In contrast to spheres, for
spheroidal particles the scattering matrix is usually given for North
in the plane of incidence before the scattering event and North in the
plane of departure after the scattering event.

In the test cases for spheroidal grains, we use spherical grains which
are treated in the computations as if they were spheroids.  Therefore,
an orientation $\bfo$ is artificially assigned to these spherical
grains. The orientation $\bfo$ can be chosen either fixed or even at
random without altering the result of the scattering computation.  In
addition, the scattering matrix of the spheres is multiplied by two
rotation matrices $\bfR$ to account for the different orientations of
the North direction during the scattering process on spheroidal
grains. These two rotations describe the change of the North direction
from the plane of incidence to the plane of scattering by the angle
$\varphi$ and the rotation from the plane of scattering to the plane
of departure by the angle $\gamma$
(Fig.~\ref{p2:fig:scattering_geometry}).  The scattering matrix
$\bfZ_\text{sph}$ of the so artificially assigned orientation of
spherical particles, therefore,
\begin{align}
\bfZ_\text{sph} &=  \bfR (\gamma)\bfZ (\theta) \bfR (\varphi)\\
 &= \begin{pmatrix}
a & b x & b y & 0\\
b p & a p x - c q y & a p y + c q x & d q \\
-b q &-a q x-c p y & -a q y + c p x & d p\\
0 & d y &-d x & c
\end{pmatrix}\label{p2:eq:sphereToSpheroid}
\end{align}
where $x = \cos2\gamma$, $y = \sin2\gamma$, $p = \cos2\varphi$, and $q
= \sin2\varphi$. The angle $\gamma$ is calculated using the angle of
incidence $\alpha$ and the scattering angles $\theta$ and $\varphi$ as
described in Appendix~\ref{p2:sec:calculatingGamma}.

\section{Results}

\subsection{Numerical set-up}

We verified the numerical treatment of the analytical test cases
    TC~1 - TC~4 as implemented in {\name}, while \citetalias{Peest2017}
    verified them using SKIRT. The MC treatments of both codes differ
    in many aspects. In SKIRT various acceleration methods such as
    forced scattering and forced absorption are included that are not
    implemented in {\name}. SKIRT \citep{CampsBaes15} uses a
    simplified Stokes formalism, which is derived assuming spherical
    particles; whereas {\name} treats spheroidal-shaped
    particles. SKIRT and {\name} use different vectorization
    technologies.  The model grids in both codes are
    different. \citetalias{Peest2017} used for the test cases a grid
    of $(601,601,60)$ cuboids each having a length of $(0.003, 0.003,
    3\times10^{-7})$ in arbitrary units and for each cube that
    includes electrons an optical depth of $\tau=10^{-3}$. {\name}
    uses a cartesian coordinate system with cubes for which any of
    these cubes can be divided into sub-cubes. We find for the test
    cases a good {\name} set-up using $(587,587,3)$ cubes with a side
    length of each cube of one.  The cube in the center and cubes of
    the slabs are filled by electrons at an optical depth of
    $\tau=0.5$, other cubes are free of electrons and have
    $\tau=0$. Hence, the probability of multiple scattering events in
    cells filled with electrons is unlikely but not zero. Photons that
    scatter a second time either at the center or at the slabs are
    ignored so that one can compare the numerical results with the
    analytical solutions. Therefore, the choice of $\tau$ shall be
    done with some care. A too-low value of $\tau$ reduces the
    scattering probability of the photons while an increase of $\tau$
    enhances the chance of multiple scattering and will lead to
    photons that must be ignored.

The number of photon packets launched is unless otherwise stated
$N_{\gamma} = 2.5 \times 10^{10}$. Results when decreasing or
increasing the number of emitted photon packets and applying different
sampling of the model space are discussed in
Sect.~\ref{precision.sec}. We repeat test cases TC~1 - TC~3 of
\citetalias{Peest2017} using the scattering matrix $\bfZ_\text{sph}$
and pretending that sphere-like grains have an orientation and
therefore, need to be treated as spheroids. We choose the grain
orientation along $+\hat{e}_z$.

\subsection{Testcases TC~1 - TC~4}

\begin{figure*}
\centering
\includegraphics[width = 2\columnwidth, trim={1.9cm -0.6cm 1.4cm -0.1cm},clip]{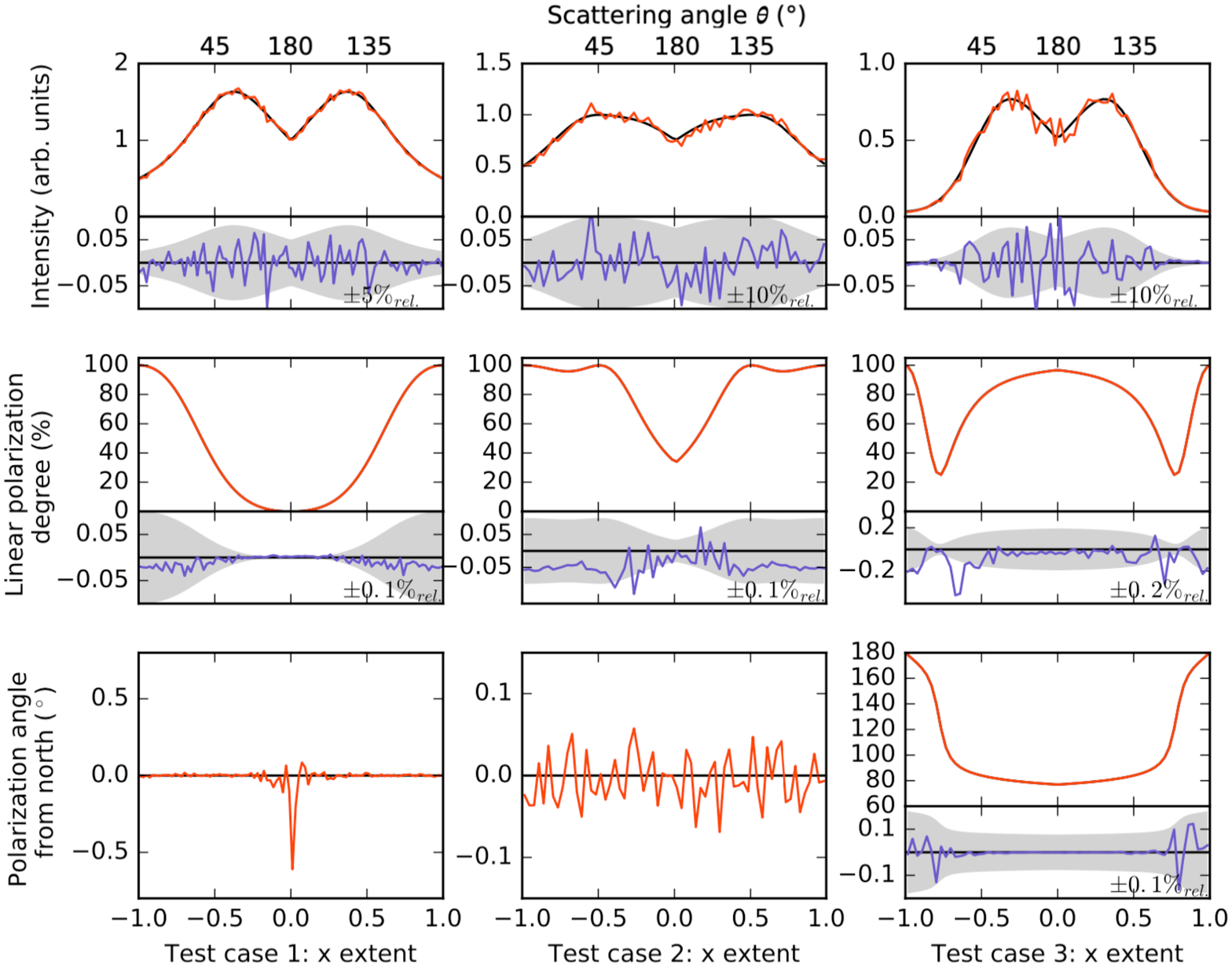}
\caption{Test cases 1 through 3 (left to right columns) using the
  spheroid-like M\"uller matrix (Eq.~\ref{p2:eq:sphereToSpheroid})
  with assumed particle orientation along the z-axis. Intensity (top
  row), linear polarization degree (middle row), and polarization
  angle (bottom row) of the observed radiation.  The panels show the
  analytical solution (black) and the model results (orange).  The
  bottom panels show the absolute differences (blue) and relative
  differences (shaded area) between the analytic solution and the
  model, the magnitude of the shaded area is given in every panel.}
\label{p2:analyticalTestcasesFM.fig}
\end{figure*}

\begin{figure}
\centering
\includegraphics[width = \columnwidth]{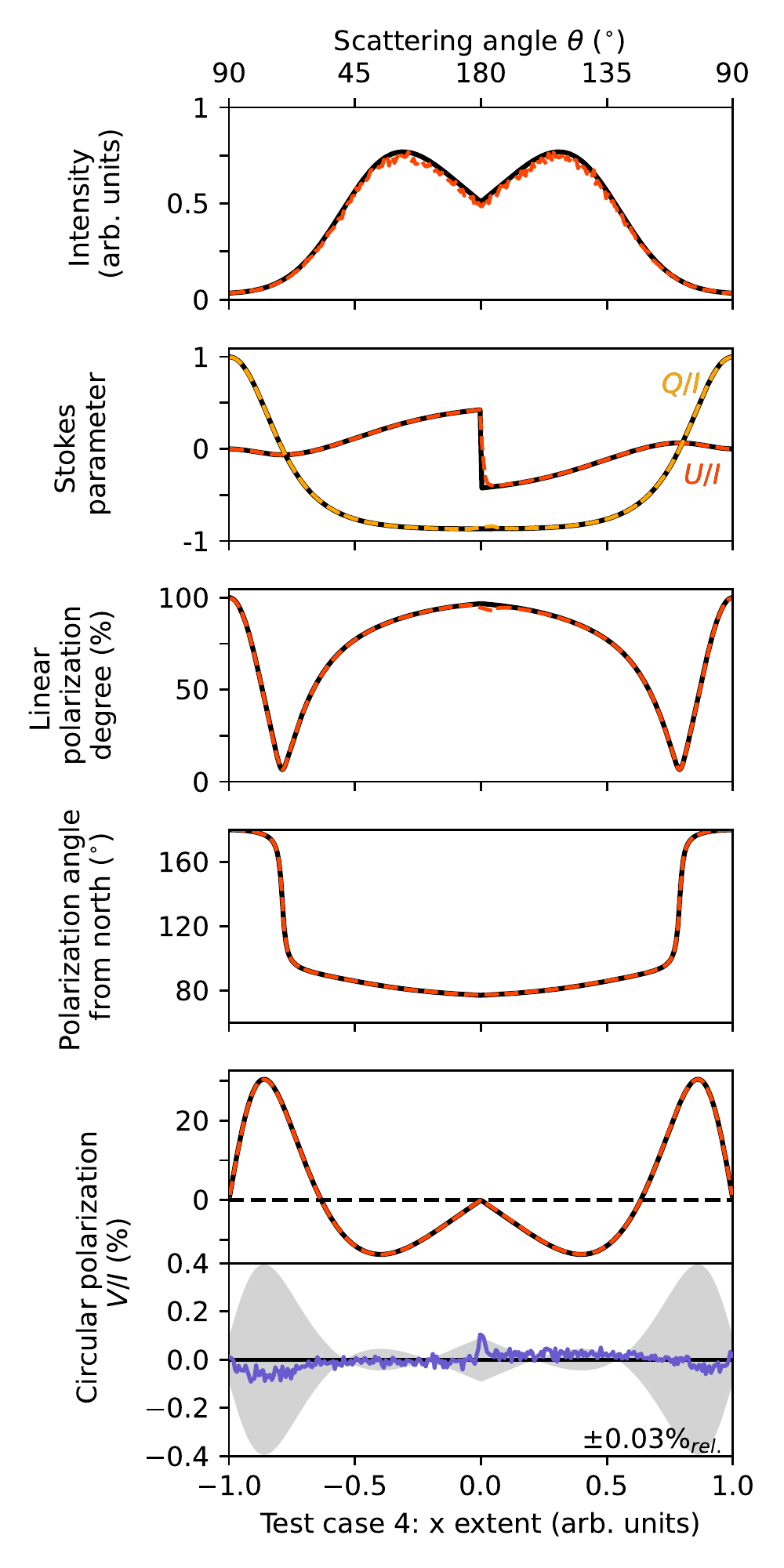}
\caption{Circular polarization test case from \citetalias{Peest2017}
  using the spheroid-like M\"uller matrix
  (Eq.~\ref{p2:eq:sphereToSpheroid}) with assumed orientation of the
  particles along the z-axis.  We show the intensity $I$, Stokes $U/I$
  and $Q/I$, the linar polarisation degree and angle, and the circular
  polarization $V/I$.  The analytic solution are indicated by the
  dashed lines in black and the {\name} model results in red.  Bottom:
  The absolute differences (blue) and relative differences (shaded
  area) between the analytic solution and the model with a typical
  relative error of $\sigma \left(\Delta(x)\right) = 3 \times 10^{-4}$
  (Eq.~\ref{delta.eq}) as labelled.}
\label{p2:TC4FM.fig}
\end{figure}

We compare the numerical results of {\name} for the test cases TC~1 -
TC~4 against the analytical solutions as provided in Eq.~44 - 63 by
\citetalias{Peest2017}. In Fig.~\ref{p2:analyticalTestcasesFM.fig} and
Fig.~\ref{p2:TC4FM.fig} the analytic solutions are shown as black
lines and the result of {\name} with red lines along the instrumental
position $x$ of the detector (Fig.~\ref{p2:fig:scattering_geometry}).
\citetalias{Peest2017} derived the scattering angle $\theta $ for a
given detector position, which we also label in
Fig.~\ref{p2:analyticalTestcasesFM.fig} and Fig.~\ref{p2:TC4FM.fig},
by

\begin{equation}
  \theta = \frac{\pi} {2} \pm {\rm atan} \left( \frac{1}{|\, x \,|} - 1 \right)
\end{equation}

The intensity profiles in arbitrary units are shown in the top panels
of (Fig.~\ref{p2:analyticalTestcasesFM.fig}).  At the bottom of each
sub-panel, the absolute deviation from the analytic solution is shown
with blue lines and the grey-shaded areas mark relative deviations
with a typical error as labelled.  The left column of
Fig.~\ref{p2:analyticalTestcasesFM.fig} shows the results of the first
test case (TC~1).  The result of {\name} jitters around the analytic
solution. The middle graph shows the linear polarization degree
computed using Eq.~3 in \citetalias{Peest2017}. {\name} reproduces the
analytic solution of TC~1 to better than $\pm 0.1\%_{ \rm {rel}}$. We
attribute residual deviations to sampling errors in the initial
direction of the photons from the source and the finite size of the
model grid.  The analytical curves are calculated for an infinitesimal
``height'' of the slabs. In the numerical treatment, the photon
packages that arrive at a given detector pixel have taken paths with a
small but finite height. In the outer parts ($\|x\|>0.3$), these paths
lead to a lower average polarization degree. In the inner parts
($\|x\|<0.3$), they lead to a higher average polarization degree.  In
the bottom graph, the polarization angle computed using Eq.~4 in
\citetalias{Peest2017} is compared to the analytic result which is
$0\degree$. Here {\name} deviates from the analytical result in the
very central region ($\|x\|<0.1$). Such inaccuracies cannot be avoided
and are expected due to the set-up of the numerical grid, which is the
finite height of the slabs, and the amplification of the noise for
polarization degrees close to zero.

The middle column of Fig.~\ref{p2:analyticalTestcasesFM.fig} shows the
result of the second test case (TC~2). The intensity curve includes
the noise of several per cent and is higher than in TC~1. This is because
the chance of scattering at the central cell is lower than unity. Some
photons do not scatter so fewer photons propagate towards the
slabs. This leads to a lower photon count at the detector. The
polarization degree for TC~2 is correct to $0.1\%_{ \rm {rel}}$
(Fig.~\ref{p2:analyticalTestcasesFM.fig}). As in the case of TC~1, we
attribute the remaining differences to the resolution of the grid. The
analytic solution of the polarization angle is zero and is shown
together with the numerical solution by {\name} at the bottom of
Fig.~\ref{p2:analyticalTestcasesFM.fig}.

The right column of Fig.~\ref{p2:analyticalTestcasesFM.fig} shows the
result of the third test case (TC~3).  The intensity curve is plotted
in the first row. The model results follow the analytic solution. The
noise in the numerical solution is comparable to the noise found in
TC~2 and with the same explanations. The polarization degree derived by
{\name} is for TC~3 correct to well below $0.2\%_{\rm {rel}}$. As we
noted in \citetalias{Peest2017}, the area around $0.6\leq\|x\|\leq0.7$
is very difficult to treat at high precision. In this area, the
scattering angle from the initial direction to the slabs changes only
by $0.1^\circ$, whereas the polarization degree by $\sim 40\%_{\rm
  {abs}}$. The scattering matrix is tabulated for integer angles and
linearly interpolated for scattering angles in between. This
comparatively sparse sampling is sufficient to calculate
the correct result. In comparison to \citetalias{Peest2017}, we even
attain a higher precision. We attribute this to the fact that the
scattering matrix for spheroids not only contains the scattering but
also, the rotations of the plane of scattering. In
\citetalias{Peest2017} we calculated the effect of the plane rotations
separately.  Variations of the polarization angle are shown in the
third row of Fig.~\ref{p2:analyticalTestcasesFM.fig}. {\name} applied
to this test case follows the analytic result to the high precision of
$0.1\%_{ \rm {rel}}$.

Scattering at electrons does not lead to circular polarization, as
$d(\theta) = 0$ in Eq.~\eqref{p2:eq:scatMatSphere}. In
\citetalias{Peest2017} we introduced a hypothetical particle for which

\begin{align}
    d(\theta)&=-\cos\theta\sin\theta\\
    c(\theta)&=\cos\theta\cos\theta
\end{align}

\noindent
otherwise, these test particles behave as electrons. Such particles
lead to circular polarization. They are applied in TC~4, in which the
geometry of TC~3 is used. In Fig.~\ref{p2:TC4FM.fig} the results of TC~4
are displayed for {\name} launching the large number of $N_{\gamma} =
2.5 \times 10^{11}$ photon packets.  The intensity, the reduced Stokes
parameters $Q/I$ and $U/I$, the linear polarization degree, the
 polarization angle and the circular polarization expressed as $V/I$ is
shown. {\name} (red) reproduces the analytic (black) solutions of
Stokes parameters of TC~4 at high precision and for the circular
polarization to better than $0.03\%_{ \rm {rel}}$.

\subsection{Numerical precision \label{precision.sec}}

The precision of {\name} is exemplified for computations of circular
polarized light by varying the number of sub-cubes $N_{\rm {sub}}$ of
cells along the $x, y$, and $z$ coordinates and the number of emitted
photons $N_{\gamma}$. The difference between the numerical model
${V}/{I}_{{\rm {\name}}}$ and the analytical solution
${V}/{I}_{{\rm{ana}}}$ (\citetalias{Peest2017}) is computed for the
reduced Stokes parameter for circular polarized light of TC~4

\begin{equation} \label{delta.eq}
  \Delta = \frac{V}{I}_{_{\rm {\name}}}- \ \ \frac{V}{I}_{{\rm{ana}}} \ .
  \end{equation}

The minimum, maximum and standard deviation $\sigma $ of that
difference $\Delta$ are given for the parameter variations in
Table~\ref{res.tab}. For the minimum number of emitted photons
($N_{\gamma}= 2.5 \times 10^{9}$) an increase of $N_{\rm {sub}}$ from
1 to 3 does not improve the $1\sigma$ noise in the model and
$N_{\gamma}$ shall be increased.  Launching 10 times more photons the
noise in the model without sub-cubes ($N_{\rm {sub}}=1$) remains
whereas it improves by a factor of 2 for the $N_{\rm {sub}}=3$ and
almost by a factor of 4 for the $N_{\rm {sub}}=11$ model. In models
without sub-cubes the location of the first scattering event, this is
the blob of electrons in the center of the model space, is not well
sampled. This causes a systematic error in backward scattering
($\theta \sim 180^{\rm o}$) at $x \sim 0$. The effect is reduced by
increasing the number of sub-cubes and for $N_{\rm {sub}}=11$
decreases to $\Delta \ (x \sim 0) < 10^{-3}$ as shown in the bottom
panel of Fig.~\ref{p2:TC4FM.fig}. Even for our choice of
optically thin cubes, the detailed fine sampling impacts the precision
that can be reached, and which cannot be improved by increasing simply
the photon statistics (Table~\ref{res.tab}).  Obviously, with an even
finer grid and more photons, the precision of {\name} could be improved
further at the expense of a boost in computer power. In our server
environment, the latter model took (already) $\sim 200$\,h. For
reaching similar precisions both codes show similar run times.

\begin{table}[!htb]
\begin{center}
  \caption {Precision of {\name} for TC~4 by varying the number of
    sub-cubes and emitted photons. The minimum, maximum and the
    standard deviation $\sigma $ of the model result minus the
    analytical solution in $V/I$ is specified using $\Delta$ as in
    Eq.~(\ref{delta.eq}) \label{res.tab}.}
\begin{tabular}{r l | r c r}
\hline\hline
 $N_{\rm {sub}}$ & $N_{\gamma}$&$\min$& $\max$ & $\sigma $ \\
               &              &    \multicolumn{3}{c}{$\ (10^{-4})$}\\
\hline
1  &  $ 2.5 \times 10^{9}$  & -35   & 28  &  18\\
1  &  $ 2.5 \times 10^{10}$ & -26   & 23  &  18\\
3  &  $ 2.5 \times 10^{9}$  & -63   & 64  &  20\\
3  &  $ 2.5 \times 10^{10}$ & -25   & 22  &   9\\
11 &  $ 2.5 \times 10^{10}$ & -16   & 13  &   5\\
11 &  $ 2.5 \times 10^{11}$ & -9    & 10  &   3\\
\hline
\end{tabular}
\end{center}
\end{table}


\subsection{Dichroism and birefringence}\label{p2:validateDichroBire.sec}

In addition to the polarization due to scattering, we have to confirm
the implementations of polarization due to dichroism and
birefringence.  For this a spherical distribution of dust with density
$n$ and radius $r$ around a central source are considered.  The test is
performed using hypothetical dust particles which dichroically absorb
and slows radiation, but which do not scatter. This is to remove any
side effects from the scattering implementation. Such hypothetical
dust particles are chosen to be analytically simple,
\begin{subequations}\label{p2:eq:dcTest1C}
\begin{align}
\cext     \cdot n \cdot r &= 2.2\\
\cpol     \cdot n \cdot r &= 2 \sin\alpha\\
\ccpol    \cdot n \cdot r &=   \cos\alpha\\
\csca     \cdot n \cdot r &= 0\\
\cscapol \cdot n \cdot r &= 0
\end{align}
\end{subequations}
The angle $\alpha$ is the angle of incidence
(Fig.~\ref{p2:fig:scattering_geometry}) and the sine and
cosine make it such that the transition is smooth for sight lines
around $\alpha = 0$. Consider initially right-handed circular
polarized radiation, $\bfS=(1,0,0,1)$, that is travelling through the
dust cloud. Its direction and the grain orientation have a constant
angle of incidence $\alpha$. Following
Eq.~\eqref{p2:eq:dichroicExtinction}, upon leaving the simulation area
the Stokes vector of the photon package will be
\begin{equation}\label{p2:eq:STest1C}
\bfS = \e{-2.2}
\begin{pmatrix}
 \cosh( 2\sin\alpha )\\
-\sinh( 2\sin\alpha )\\
-\sin(   \cos\alpha )\\
 \cos(   \cos\alpha )
\end{pmatrix}
\end{equation}

Note that in this scenario the Stokes vector depends only on $\alpha$.
The analytic solutions for the reduced Stokes parameters are,
\begin{subequations}\label{p2:eq:dcTest1ana}
\begin{align}
I &= \e{-2.2}\cosh( 2\sin\alpha )\\
Q/I &= -\tanh( 2\sin\alpha )\\
U/I &= \frac{-\sin(   \cos\alpha )}{\cosh( 2\sin\alpha )}\\
V/I &= \frac{\cos(   \cos\alpha )}{\cosh( 2\sin\alpha )}
\end{align}
\end{subequations}
under the different viewing angles $\alpha$ towards the z-axis. We use
this closed form validates the implementation of dichroism and
birefringence in {\name}.


\begin{figure}
\centering
\includegraphics[width =\columnwidth]{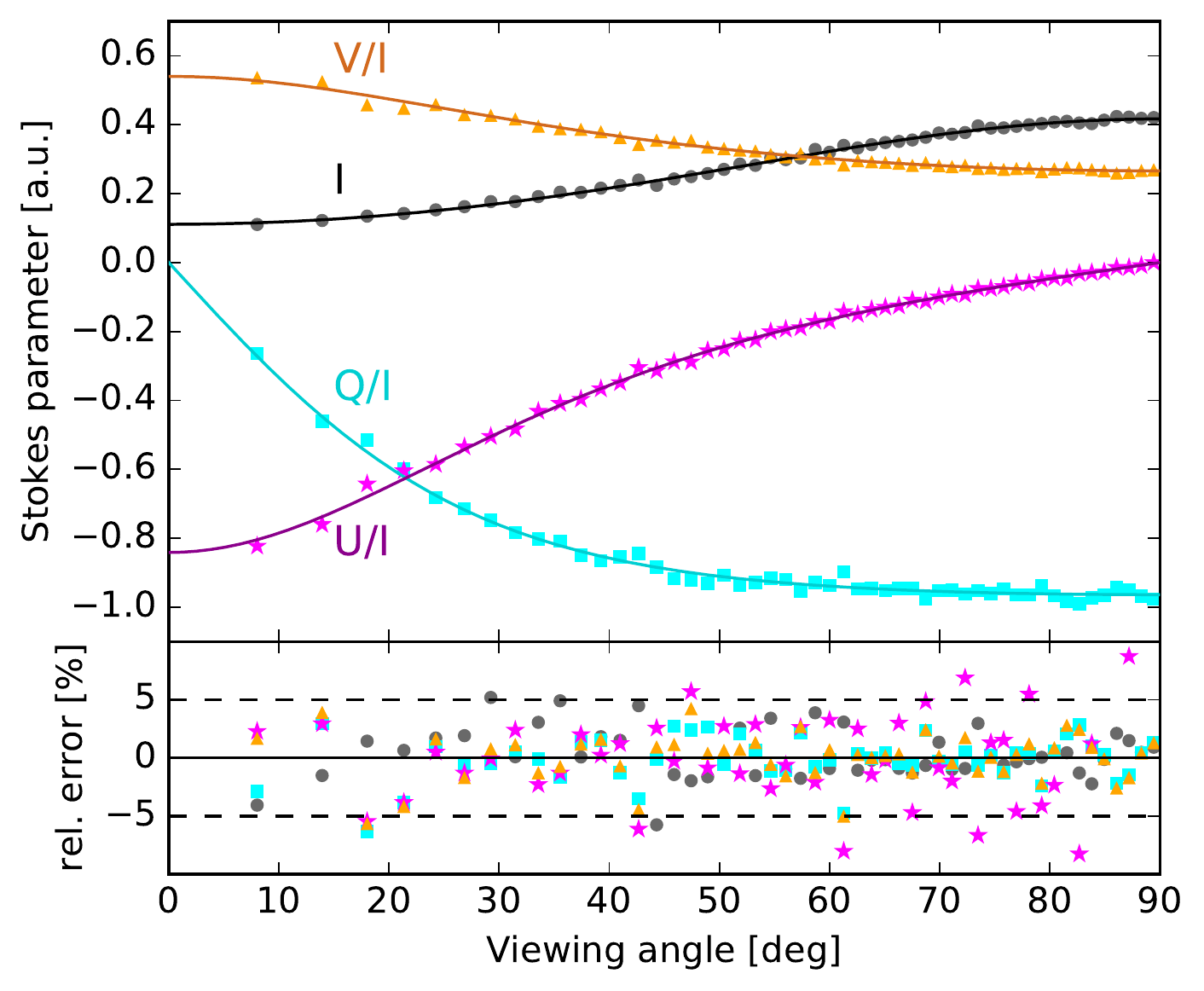}
\caption{Dichroism and birefringence test case.  Intensity $I$
  (black), reduced Stokes $Q/I$ (cyan), $U/I$ (magenta), and circular
  polarization $V/I$ (orange) for different viewing angles.  Analytic
  solutions (Eq.~\ref{p2:eq:dcTest1ana}) are represented by lines and
  {\name} results by symbols.  The relative error of the numerical run
  against the analytic solution is shown in the bottom panel using
  the same color code. }
\label{p2:dcTest1.fig}
\end{figure}


In Fig.~\ref{p2:dcTest1.fig} we compare the results of {\name} with the
analytic solutions.  In the top panel, the observed Stokes parameters
are plotted, using lines for the analytic solution and symbols for the
numerical results achieved by {\name}.  The lower panel gives the
relative error between the numerical and analytical solutions.  As we
bin the exiting photons using the cosine of their exit angles, there
are more data points for large viewing angles than for small viewing
angles.  Generally, the analytic solution is reproduced to better than
$5\%_{ \rm {rel}}$. There are a few aliasing effects at $18^\circ$, $33^\circ$,
$42^\circ$ and $61^\circ$ $U/I$ shows a bit larger noise. We attribute this to
sampling errors caused by the finite resolution of the grid.


\subsection{Albedo test case}

The final test confirms the correct implementation of the albedo. The
albedo of spheroids and spheroid-like particles depends on the
polarization of the radiation interacting with the grains. {\name}
handles this using Eq.~\eqref{p2:eq:albedoDichroism}. The albedo
connects the effects of scattering and extinction, and a test of the
implementation needs to consider both effects simultaneously.

\begin{figure} [!htb]
\centering
\includegraphics[width =0.8\columnwidth]{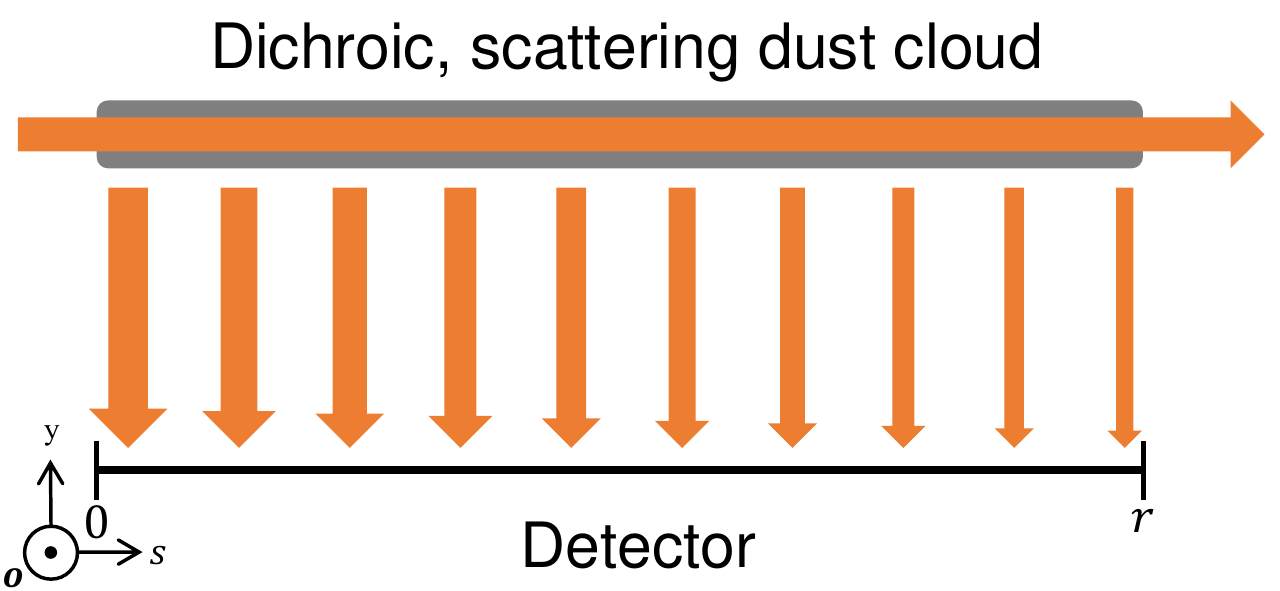}
\caption{Geometry of the albedo test case. A collimated beam passes
  along $s$ from $0$ to $r$ through a dichroic dust cloud going and
  some of the photons are scattered to a detector. The alignment of
  the dust grains $\bfo$ is towards the reader, so perpendicular to
  the beam, $\alpha = 90^\circ$.}
\label{p2:albedoTestcase.fig}
\end{figure}

We set up a test in which a collimated beam of right-handed circular
polarized radiation propagates along an elongated dust cloud of length
$r$. The dust cloud dichroically extinct and scatters the
radiation. The orientation $\bfo$ of the dust particles is
perpendicular to the beam ($\alpha = 90^\circ$,
Fig.~\ref{p2:fig:scattering_geometry}) and no radiation scatters into
the beam. The geometry of the albedo test case is illustrated in
Fig.~\ref{p2:albedoTestcase.fig}.  We use an artificial dust grain
that scatters isotropic while preserving the Stokes parameters. This
is obtained by using the 4D unity matrix as the scattering matrix,
\begin{equation}
\bfZ_\text{iso}(\lambda,\alpha,\varphi,\theta) = \mathbb{1}_4 \ .
\end{equation}
The cross-sections are similar to the previous test
(Eq.~\ref{p2:eq:dcTest1C}), but this time with the scattering cross
sections differing from zero:

\begin{subequations}\label{p2:eq:dcTest2C}
\begin{align}
\csca     \cdot n \cdot r &= 0.1\\
\cscapol \cdot n \cdot r &= -0.1
\end{align}
\end{subequations}
The Stokes vector along $0 \leq s \leq r$ in radial direction $r$ of
the dust cloud before scattering is
\begin{equation}
\bfS (s) = \e{-2.2 s/r}
\begin{pmatrix}
 \cosh( 2 s/r )\\
-\sinh( 2 s/r )\\
0\\
1
\end{pmatrix}\ .
\end{equation}


As the beam is collimated, the inverse square law does not apply
here. The scattering probability is given by
Eq.~\eqref{p2:eq:scaCrosSec},
\begin{equation}\label{p2:eq}
\tilde{C}_\text{sca} = 0.1 (1 + \tanh(2 s/r) )\ .
\end{equation}

The Stokes components of the radiation arriving at the detector are
given by multiplying $\tilde{C}_\text{sca}$ with $\bfS (s)$,

\begin{subequations}\label{p2:eq:dcTest2ana}
\begin{align}
I &= 0.1\e{-0.2\ s/r}\\
Q/I &= -\tanh( 2\ s/r) \label{eq.albedoQI} \\
U/I &= 0  \label{eq.albedoUI} \\
V/I &= 1/\cosh( 2\ s/r ) \label{eq.albedoVI}
\end{align}
\end{subequations}

The change of the Stokes components can be illustrated as follows:
The component of the radiation that is polarized parallel to the
orientation $\bfo$ of the grains (out of the plane of the paper in
Fig.~\ref{p2:albedoTestcase.fig}) is extinct stronger than the component
perpendicular to $\bfo$. Along the path through the dust cloud, the
remaining radiation is more polarized in the plane of the paper, hence
$Q < 0$.
Following Eq.~\ref{eq.albedoVI} the circular polarization degree $V/I$
decreases along $s$ (Fig.~\ref{p2:albedoTestcase.fig}). As the
incidence of the radiation is perpendicular to the alignment
($\alpha=\pi/2$), the phase between the two components stays constant
($\ccpol=0$) and the circular polarization does not turn into linear
polarization ($U\equiv0$).


\begin{figure}[htb]
\centering
\includegraphics[width = \columnwidth]{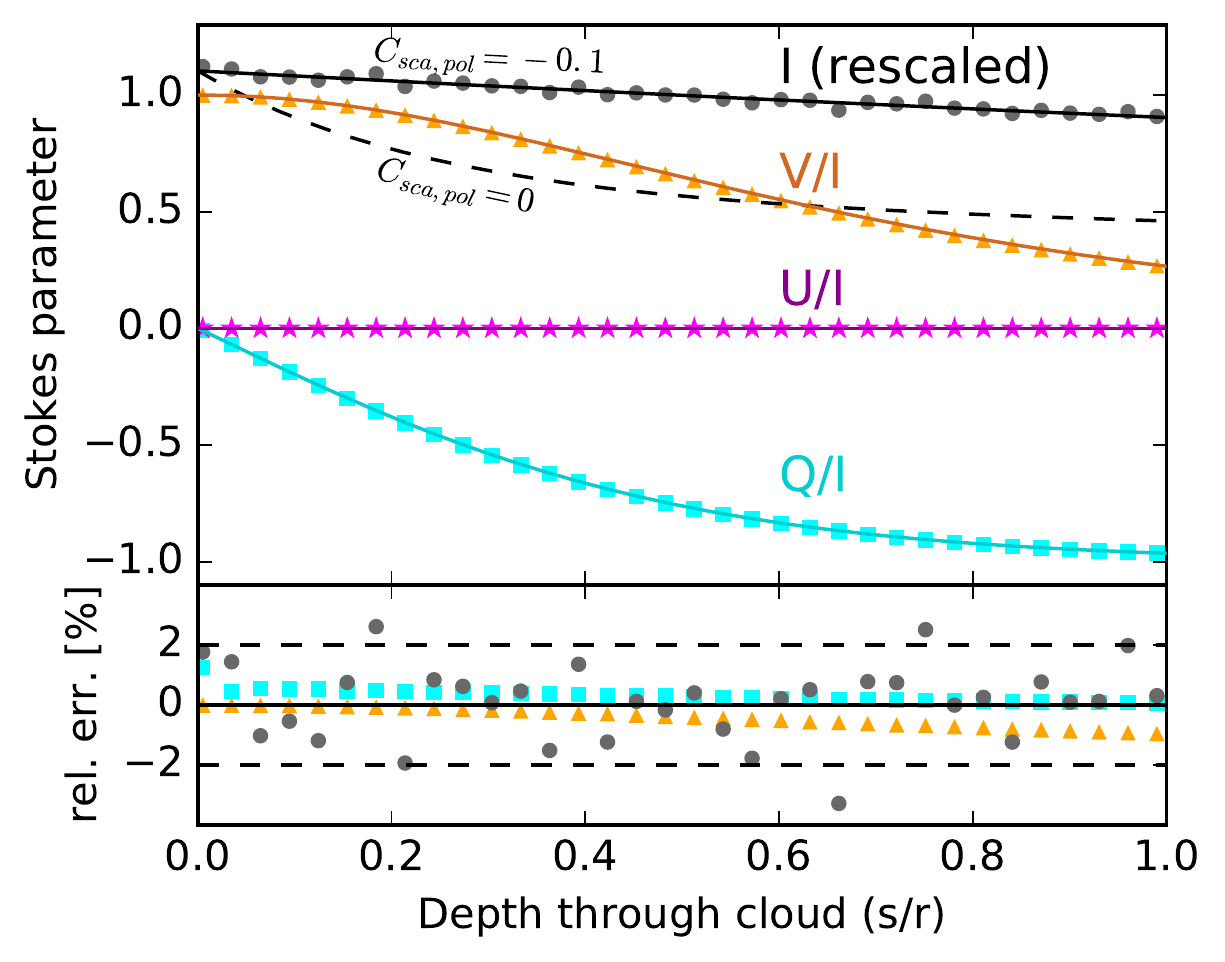}
\caption{Albedo test case.  Elements of the reduced Stokes vector of
  photons scattered out of a cloud having spheroidal particles with
  $\cscapol=-0.1$.  Intensity using spheroids with $\cscapol= 0$ is
  shown as a dashed line.  Analytic solutions
  (Eq.~\ref{p2:eq:dcTest2ana}) are represented by lines and {\name}
  results by symbols. Intensities in black are multiplied by factor 11;
  rest of notation as in Fig.~\ref{p2:dcTest1.fig} and Table~\ref{sec.notation}. }
\label{p2:dcTest2.fig}
\end{figure}


In Fig.~\ref{p2:dcTest2.fig} we show the results of the albedo test
case solved by {\name} along with the analytic solutions. The
intensity reduces by $\sim 20$\,\% along the profile of the dust
cloud. The circular polarization degree ($V/I$) of the radiation
reduces from 1 to about 0.27. These effects follow the analytic
expectation with high precision, the relative error is below 0.3\%. In
summary, {\name} reproduces the correct behavior of the analytic
solution of the albedo test case and the numerical imprecision of the
Stokes vector is low ($\simless 1$\,\%).

If the albedo would not favor reflecting radiation
polarized perpendicular to the alignment, the reflected intensity
would behave very differently. As an example, we show the analytic
solution of the intensity for test particles having $\cscapol = 0$. In
that case, the total intensity of the beam decreases exponentially as
often applied in astronomy.


\section{Summary}\label{p2:sec:Conclusion}
We have implemented polarization of radiation by spheroidal dust
considering scattering, dichroic extinction, and birefringence in the
Monte Carlo radiative transfer code {\name}. We provide a detailed
description of the methodology used, for which the Stokes formalism
was selected.  We developed paradigmatic examples for testing such
numerical polarization radiative transfer models using spheroidal dust
against analytical solutions.  They are used to verify the correct
functionality of {\name} and they may be applied by other teams to
verify their codes. Further, the comparison of {\name} against the
analytical solutions provides means for estimating the numerical
precision. We find a typical deviation of {\name} from the analytical
solution for the intensity, the linear polarization degree and angle
of 0.1\% and in the circular polarization of 0.03\%.  However,
situations, e.g., at a small angle of incidence $\alpha \sim 0$ or for
backward scattering ($\theta \sim 180^{\rm o}$), we notice larger
deviations that are due to sampling errors caused by the finite size
of the applied grid of the model volume.

The analytical test cases are suited to other teams for estimating the
precision of their dust polarization treatments. So far, only a
limited number of MCRT codes support polarization including more
complicated particle geometries than spheres. This
is even though extinction due to spheres cannot explain the well-established
Serkowski curve \citep{Serkowski1975} of the diffuse ISM. We hope that
the presented examples with their attached analytical solutions will
help expand the number and ease the verification of MCRT codes that
treat dust polarization of more complex grain shapes.

In realistic applications, one needs to consider more appropriate dust
geometries, e.g., replacing spherical with spheroidal grain shapes as a
starting point. For a mix of such particles that are of different kind
of carbon and silicate materials and which span a range of particle
sizes and grain porosity, their optical cross sections and the
scattering matrix needs to be computed, furthermore some kind of grain
alignment need to be implemented. {\name} is capable of tracing the
polarization due to spheroidal dust grains.  The here presented and
tested code is open to the community in a collaborative effort.

\begin{acknowledgements}
C.P. and M.B. acknowledge the financial support from CHARM
(Contemporary physical challenges in Heliospheric and Astrophysical
Models), a Phase-VII Inter-university Attraction Pole program
organized by BELSPO, the Belgian federal Science Policy Office.
\end{acknowledgements}
\bibliographystyle{aa} 
\bibliography{Library}

\begin{thebibliography}{69}
\expandafter\ifx\csname natexlab\endcsname\relax\def\natexlab#1{#1}\fi

\bibitem[{{Abhyankar} \& {Fymat}(1969)}]{Abhyankar1969}
{Abhyankar}, K.~D. \& {Fymat}, A.~L. 1969, Journal of Mathematical Physics, 10,
  1935

\bibitem[{{Asano} \& {Yamamoto}(1975)}]{Asano1975}
{Asano}, S. \& {Yamamoto}, G. 1975, \ao, 14, 29

\bibitem[{{Baes} \& {Camps}(2015)}]{Baes2015}
{Baes}, M. \& {Camps}, P. 2015, Astronomy and Computing, 12, 33

\bibitem[{{Baes} {et~al.}(2019){Baes}, {Peest}, {Camps}, \&
  {Siebenmorgen}}]{Baes19}
{Baes}, M., {Peest}, C., {Camps}, P., \& {Siebenmorgen}, R. 2019, \aap, 630,
  A61

\bibitem[{{Bertrang} \& {Wolf}(2017)}]{Bertrang2017}
{Bertrang}, G.~H.-M. \& {Wolf}, S. 2017, \mnras, 469, 2869

\bibitem[{{Bianchi} {et~al.}(1996){Bianchi}, {Ferrara}, \&
  {Giovanardi}}]{Bianchi1996}
{Bianchi}, S., {Ferrara}, A., \& {Giovanardi}, C. 1996, \apj, 465, 127

\bibitem[{Bohren \& Huffman(1998)}]{Bohren1998}
Bohren, C.~F. \& Huffman, D.~R. 1998, Absorption and scattering of light by
  small particles (New York (N.Y.): Wiley-Interscience)

\bibitem[{{Camps} \& {Baes}(2015)}]{CampsBaes15}
{Camps}, P. \& {Baes}, M. 2015, Astronomy and Computing, 9, 20

\bibitem[{{Camps} \& {Baes}(2020)}]{Camps2020}
{Camps}, P. \& {Baes}, M. 2020, Astronomy and Computing, 31, 100381

\bibitem[{{Chandrasekhar}(1960)}]{Chandrasekhar1960}
{Chandrasekhar}, S. 1960, {Radiative transfer} (New York: Dover)

\bibitem[{Cheng \& Gupta(1989)}]{Cheng1989}
Cheng, H. \& Gupta, K. 1989, Journal of Applied Mechanics, 56, 139

\bibitem[{{Chira} {et~al.}(2016){Chira}, {Siebenmorgen}, {Henning}, \&
  {Kainulainen}}]{Chira2016}
{Chira}, R.~A., {Siebenmorgen}, R., {Henning}, T., \& {Kainulainen}, J. 2016,
  \aap, 592

\bibitem[{Contopoulos \& Jappel(1974)}]{IAU1974}
Contopoulos, G. \& Jappel, A., eds. 1974, {Transactions of the IAU}, Vol. 15B
  (Springer Netherlands), 166

\bibitem[{Devroye(2013)}]{Devroye2013}
Devroye, L. 2013, Non-Uniform Random Variate Generation (Springer New York)

\bibitem[{{Draine}(1988)}]{Draine1988}
{Draine}, B.~T. 1988, \apj, 333, 848

\bibitem[{{Draine} \& {Flatau}(1994)}]{Draine1994}
{Draine}, B.~T. \& {Flatau}, P.~J. 1994, Journal of the Optical Society of
  America A, 11, 1491

\bibitem[{{Draine} \& {Hensley}(2021)}]{DH21}
{Draine}, B.~T. \& {Hensley}, B.~S. 2021, \apj, 919, 65

\bibitem[{{Fleck} \& {Canfield}(1984)}]{Fleck1984}
{Fleck}, Jr., J.~A. \& {Canfield}, E.~H. 1984, Journal of Computational
  Physics, 54, 508

\bibitem[{{Forrest} {et~al.}(1975){Forrest}, {Gillett}, \&
  {Stein}}]{Forrest1975}
{Forrest}, W.~J., {Gillett}, F.~C., \& {Stein}, W.~A. 1975, \apj, 195, 423

\bibitem[{{Goosmann} \& {Gaskell}(2007)}]{Goosmann2007}
{Goosmann}, R.~W. \& {Gaskell}, C.~M. 2007, \aap, 465, 129

\bibitem[{{Goosmann} {et~al.}(2014){Goosmann}, {Gaskell}, \&
  {Marin}}]{Goosmann2014}
{Goosmann}, R.~W., {Gaskell}, C.~M., \& {Marin}, F. 2014, Advances in Space
  Research, 54, 1341

\bibitem[{{Gordon} {et~al.}(2017){Gordon}, {Baes}, {Bianchi}, {Camps},
  {Juvela}, {Kuiper}, {Lunttila}, {Misselt}, {Natale}, {Robitaille}, \&
  {Steinacker}}]{Gordon2017}
{Gordon}, K.~D., {Baes}, M., {Bianchi}, S., {et~al.} 2017, \aap, 603, A114

\bibitem[{Hamaker \& Bregman(1996)}]{Hamaker1996}
Hamaker, J. \& Bregman, J. 1996, Astronomy and Astrophysics Supplement Series,
  117, 161

\bibitem[{Harries(2000)}]{Harries2000}
Harries, T.~J. 2000, Monthly Notices of the Royal Astronomical Society, 315,
  722

\bibitem[{{Heymann} \& {Siebenmorgen}(2012)}]{Heymann2012}
{Heymann}, F. \& {Siebenmorgen}, R. 2012, \apj, 751, 27

\bibitem[{{Ivezic} {et~al.}(1997){Ivezic}, {Groenewegen}, {Men'shchikov}, \&
  {Szczerba}}]{Ivezic1997}
{Ivezic}, Z., {Groenewegen}, M.~A.~T., {Men'shchikov}, A., \& {Szczerba}, R.
  1997, \mnras, 291, 121

\bibitem[{Kataoka {et~al.}(2015)Kataoka, Muto, Momose, Tsukagoshi, Fukagawa,
  Shibai, Hanawa, Murakawa, \& Dullemond}]{Kataoka2015}
Kataoka, A., Muto, T., Momose, M., {et~al.} 2015, The Astrophysical Journal,
  809, 78

\bibitem[{{Kr{\"u}gel}(2008)}]{K08}
{Kr{\"u}gel}, E. 2008, {An introduction to the physics of interstellar dust}
  (IOP)

\bibitem[{{Kr{\"u}gel}(2009)}]{Kruegel2009}
{Kr{\"u}gel}, E. 2009, \aap, 493, 385

\bibitem[{{Kr{\"u}gel}(2015)}]{Kruegel2015}
{Kr{\"u}gel}, E. 2015, \aap, 574

\bibitem[{{Lee} {et~al.}(2008){Lee}, {Seon}, {Min}, {Park}, {Yuk}, {Edelstein},
  {Korpela}, {Sankrit}, {Park}, \& {Ryu}}]{Lee2008}
{Lee}, D.~H., {Seon}, K.~I., {Min}, K.~W., {et~al.} 2008, \apj, 686, 1155

\bibitem[{Lucas(2003)}]{Lucas2003}
Lucas, P. 2003, Journal of Quantitative Spectroscopy and Radiative Transfer,
  79, 921, electromagnetic and Light Scattering by Non-Spherical Particles

\bibitem[{{Lucy}(1999)}]{Lucy1999}
{Lucy}, L.~B. 1999, \aap, 344, 282

\bibitem[{Martin(1974)}]{Martin1974}
Martin, P. 1974, The Astrophysical Journal, 187, 461

\bibitem[{{Miller} \& {Goodrich}(1990)}]{Miller1990}
{Miller}, J.~S. \& {Goodrich}, R.~W. 1990, \apj, 355, 456

\bibitem[{{Min} {et~al.}(2009){Min}, {Dullemond}, {Dominik}, {de Koter}, \&
  {Hovenier}}]{Min2009}
{Min}, M., {Dullemond}, C.~P., {Dominik}, C., {de Koter}, A., \& {Hovenier},
  J.~W. 2009, \aap, 497, 155

\bibitem[{Mishchenko(1991)}]{Mishchenko1991}
Mishchenko, M.~I. 1991, The Astrophysical Journal, 367, 561

\bibitem[{{Mishchenko}(1991)}]{Mishchenko1991b}
{Mishchenko}, M.~I. 1991, Journal of the Optical Society of America A, 8, 871

\bibitem[{Mishchenko {et~al.}(2002)Mishchenko, Travis, \&
  Lacis}]{Mishchenko2002}
Mishchenko, M.~I., Travis, L.~D., \& Lacis, A.~A. 2002, {Scattering,
  absorption, and emission of light by small particles} (Cambridge university
  press)

\bibitem[{{Mishchenko} {et~al.}(1996){Mishchenko}, {Travis}, \&
  {Mackowski}}]{Mishchenko1996}
{Mishchenko}, M.~I., {Travis}, L.~D., \& {Mackowski}, D.~W. 1996, Journal of
  Quantitative Spectroscopy and Radiative Transfer, 55, 535

\bibitem[{Montgomery \& Clemens(2014)}]{Montgomery2014}
Montgomery, J.~D. \& Clemens, D.~P. 2014, The Astrophysical Journal, 786, 41

\bibitem[{Pascucci {et~al.}(2004)Pascucci, Wolf, Steinacker, Dullemond,
  Henning, Niccolini, Woitke, \& Lopez}]{Pascucci2004}
Pascucci, I., Wolf, S., Steinacker, J., {et~al.} 2004, Astronomy \&
  Astrophysics, 417, 793

\bibitem[{Peest {et~al.}(2017)Peest, Camps, Stalevski, Baes, \&
  Siebenmorgen}]{Peest2017}
Peest, C., Camps, P., Stalevski, M., Baes, M., \& Siebenmorgen, R. 2017,
  Astronomy \& Astrophysics, 601, A92

\bibitem[{Pinte {et~al.}(2009)Pinte, Harries, Min, Watson, Dullemond, Woitke,
  M{\'e}nard, \& Dur{\'a}n-Rojas}]{Pinte2009}
Pinte, C., Harries, T., Min, M., {et~al.} 2009, Astronomy \& Astrophysics, 498,
  967

\bibitem[{{Pinte} {et~al.}(2006){Pinte}, {M{\'e}nard}, {Duch{\^e}ne}, \&
  {Bastien}}]{Pinte2006}
{Pinte}, C., {M{\'e}nard}, F., {Duch{\^e}ne}, G., \& {Bastien}, P. 2006, \aap,
  459, 797

\bibitem[{{Planck Collaboration} {et~al.}(2015){Planck Collaboration}, {Ade},
  {Aghanim}, {Alina}, {Alves}, {Armitage-Caplan}, {Arnaud}, {Arzoumanian},
  {Ashdown}, {Atrio-Barandela}, {Aumont}, {Baccigalupi}, {Banday}, {Barreiro},
  {Battaner}, {Benabed}, {Benoit-L{\'e}vy}, {Bernard}, {Bersanelli},
  {Bielewicz}, {Bock}, {Bond}, {Borrill}, {Bouchet}, {Boulanger}, {Bracco},
  {Burigana}, {Butler}, {Cardoso}, {Catalano}, {Chamballu}, {Chary}, {Chiang},
  {Christensen}, {Colombi}, {Colombo}, {Combet}, {Couchot}, {Coulais}, {Crill},
  {Curto}, {Cuttaia}, {Danese}, {Davies}, {Davis}, {de Bernardis}, {de Gouveia
  Dal Pino}, {de Rosa}, {de Zotti}, {Delabrouille}, {D{\'e}sert}, {Dickinson},
  {Diego}, {Donzelli}, {Dor{\'e}}, {Douspis}, {Dunkley}, {Dupac}, {Efstathiou},
  {En{\ss}lin}, {Eriksen}, {Falgarone}, {Ferri{\`e}re}, {Finelli}, {Forni},
  {Frailis}, {Fraisse}, {Franceschi}, {Galeotta}, {Ganga}, {Ghosh}, {Giard},
  {Giraud-H{\'e}raud}, {Gonz{\'a}lez-Nuevo}, {G{\'o}rski}, {Gregorio},
  {Gruppuso}, {Guillet}, {Hansen}, {Harrison}, {Helou},
  {Hern{\'a}ndez-Monteagudo}, {Hildebrandt}, {Hivon}, {Hobson}, {Holmes},
  {Hornstrup}, {Huffenberger}, {Jaffe}, {Jaffe}, {Jones}, {Juvela},
  {Keih{\"a}nen}, {Keskitalo}, {Kisner}, {Kneissl}, {Knoche}, {Kunz},
  {Kurki-Suonio}, {Lagache}, {L{\"a}hteenm{\"a}ki}, {Lamarre}, {Lasenby},
  {Lawrence}, {Leahy}, {Leonardi}, {Levrier}, {Liguori}, {Lilje},
  {Linden-V{\o}rnle}, {L{\'o}pez-Caniego}, {Lubin}, {Mac{\'\i}as-P{\'e}rez},
  {Maffei}, {Magalh{\~a}es}, {Maino}, {Mandolesi}, {Maris}, {Marshall},
  {Martin}, {Mart{\'\i}nez-Gonz{\'a}lez}, {Masi}, {Matarrese}, {Mazzotta},
  {Melchiorri}, {Mendes}, {Mennella}, {Migliaccio}, {Miville-Desch{\^e}nes},
  {Moneti}, {Montier}, {Morgante}, {Mortlock}, {Munshi}, {Murphy}, {Naselsky},
  {Nati}, {Natoli}, {Netterfield}, {Noviello}, {Novikov}, {Novikov},
  {Oxborrow}, {Pagano}, {Pajot}, {Paladini}, {Paoletti}, {Pasian}, {Pearson},
  {Perdereau}, {Perotto}, {Perrotta}, {Piacentini}, {Piat}, {Pietrobon},
  {Plaszczynski}, {Poidevin}, {Pointecouteau}, {Polenta}, {Popa}, {Pratt},
  {Prunet}, {Puget}, {Rachen}, {Reach}, {Rebolo}, {Reinecke}, {Remazeilles},
  {Renault}, {Ricciardi}, {Riller}, {Ristorcelli}, {Rocha}, {Rosset},
  {Roudier}, {Rubi{\~n}o-Mart{\'\i}n}, {Rusholme}, {Sandri}, {Savini}, {Scott},
  {Spencer}, {Stolyarov}, {Stompor}, {Sudiwala}, {Sutton}, {Suur-Uski},
  {Sygnet}, {Tauber}, {Terenzi}, {Toffolatti}, {Tomasi}, {Tristram}, {Tucci},
  {Umana}, {Valenziano}, {Valiviita}, {Van Tent}, {Vielva}, {Villa}, {Wade},
  {Wandelt}, {Zacchei}, \& {Zonca}}]{Ade15}
{Planck Collaboration}, {Ade}, P.~A.~R., {Aghanim}, N., {et~al.} 2015, \aap,
  576, A104

\bibitem[{{Purcell} \& {Pennypacker}(1973)}]{Purcell1973}
{Purcell}, E.~M. \& {Pennypacker}, C.~R. 1973, \apj, 186, 705

\bibitem[{{Reissl} {et~al.}(2016){Reissl}, {Wolf}, \& {Brauer}}]{Reissl2016}
{Reissl}, S., {Wolf}, S., \& {Brauer}, R. 2016, \aap, 593, A87

\bibitem[{{Robitaille}(2011)}]{Robitaille2011}
{Robitaille}, T.~P. 2011, \aap, 536, A79

\bibitem[{{Scicluna} \& {Siebenmorgen}(2015)}]{Scicluna2015b}
{Scicluna}, P. \& {Siebenmorgen}, R. 2015, \aap, 584

\bibitem[{{Seon}(2018)}]{Seon2018}
{Seon}, K.-I. 2018, ArXiv e-prints

\bibitem[{{Serkowski} {et~al.}(1975){Serkowski}, {Mathewson}, \&
  {Ford}}]{Serkowski1975}
{Serkowski}, K., {Mathewson}, D.~S., \& {Ford}, V.~L. 1975, \apj, 196, 261

\bibitem[{{Siebenmorgen}(2022)}]{S22}
{Siebenmorgen}, R. 2022, arXiv e-prints, arXiv:2211.10146

\bibitem[{{Siebenmorgen} \& {Heymann}(2012)}]{Siebenmorgen2012}
{Siebenmorgen}, R. \& {Heymann}, F. 2012, \aap, 539

\bibitem[{{Siebenmorgen} \& {Peest}(2019)}]{S19}
{Siebenmorgen}, R. \& {Peest}, C. 2019, in Astrophysics and Space Science
  Library, Vol. 460, Astronomical Polarisation from the Infrared to Gamma Rays,
  ed. R.~{Mignani}, A.~{Shearer}, A.~{S{\l}owikowska}, \& S.~{Zane}, 197

\bibitem[{{Steinacker} {et~al.}(2013){Steinacker}, {Baes}, \&
  {Gordon}}]{Steinacker2013}
{Steinacker}, J., {Baes}, M., \& {Gordon}, K.~D. 2013, Annual Review of
  Astronomy and Astrophysics, 51, 63

\bibitem[{Stokes(1852)}]{Stokes1852}
Stokes, G.~G. 1852, Transactions of the Cambridge Philosophical Society, 9, 399

\bibitem[{{Tran} {et~al.}(1997){Tran}, {Filippenko}, {Schmidt}, {Bjorkman},
  {Jannuzi}, \& {Smith}}]{Tran1997}
{Tran}, H.~D., {Filippenko}, A.~V., {Schmidt}, G.~D., {et~al.} 1997,
  Publications of the Astronomical Society of the Pacific, 109, 489

\bibitem[{Van De~Hulst(1957)}]{VanDeHulst1957}
Van De~Hulst, H.~C. 1957, Light scattering by small particles (Courier
  Corporation)

\bibitem[{{Vandenbroucke} {et~al.}(2020){Vandenbroucke}, {Baes}, \&
  {Camps}}]{Vandenbroucke2020}
{Vandenbroucke}, B., {Baes}, M., \& {Camps}, P. 2020, \aj, 160, 55

\bibitem[{{Vandenbroucke} {et~al.}(2021){Vandenbroucke}, {Baes}, {Camps},
  {Kapoor}, {Barrientos}, \& {Bernard}}]{Vandenbroucke2021}
{Vandenbroucke}, B., {Baes}, M., {Camps}, P., {et~al.} 2021, \aap, 653, A34

\bibitem[{Von~Neumann(1951)}]{vonNeumann1951}
Von~Neumann, J. 1951, Appl. Math Ser, 12, 3

\bibitem[{Voshchinnikov(2012)}]{Voshchinnikov2012}
Voshchinnikov, N. 2012, Journal of Quantitative Spectroscopy and Radiative
  Transfer, 113, 2334

\bibitem[{Voshchinnikov \& Farafonov(1993)}]{Voshchinnikov1993}
Voshchinnikov, N. \& Farafonov, V. 1993, Astrophysics and Space Science, 204,
  19

\bibitem[{Voshchinnikov \& Karjukin(1994)}]{Voshchinnikov1994}
Voshchinnikov, N. \& Karjukin, V. 1994, Astronomy and Astrophysics, 288, 883

\bibitem[{{Watson} \& {Henney}(2001)}]{Watson2001}
{Watson}, A.~M. \& {Henney}, W.~J. 2001, \rmxaa, 37, 221

\bibitem[{Whitney \& Wolff(2002)}]{Whitney2002}
Whitney, B.~A. \& Wolff, M.~J. 2002, The Astrophysical Journal, 574, 205

\bibitem[{{Witt}(1977)}]{Witt1977}
{Witt}, A.~N. 1977, \apjs, 35, 1

\bibitem[{{Yusef-Zadeh} {et~al.}(1984){Yusef-Zadeh}, {Morris}, \&
  {White}}]{Yusef-Zadeh1984}
{Yusef-Zadeh}, F., {Morris}, M., \& {White}, R.~L. 1984, \apj, 278, 186

\end{thebibliography}

\appendix

\section{Calculation of the exit angle}\label{p2:sec:calculatingGamma}

We begin with the vectors of the symmetry axis of the grain $\bfo$,
the propagation direction before scattering ${\bf k}$ and the normal
to the plane of incidence $\bfn$. By rotating $\bfn$ around $\bfk$ by
$\varphi$ we obtain the normal to the scattering plane
$\bfn_\text{scat}$. Rotating $\bfk$ around $\bfn_\text{scat}$ by
$\theta$ results in the propagation direction after scattering
$\bfk'$. The (as of yet unknown) exit angle $\gamma$ is used to rotate
$\bfn_\text{scat}$ around $\bfk'$ into the normal of the plane of
departure $\bfn'$. These rotations can be calculated with Euler's
finite rotation formula \citep{Cheng1989} and simplified by taking
into account that many pairs of these vectors are perpendicular. With
the substitutions $c_x = \cos x$ and $s_x = \sin x$ we can write
\begin{equation}
\bfn'=\bfn( c_\varphi c_\gamma- s_\varphi c_\theta s_\gamma) + \bfk\times\bfn(s_\varphi c_\gamma+ c_\theta c_\varphi s_\gamma) +\bfk\,s_\theta s_\gamma
\end{equation}
This normal must be perpendicular to the symmetry axis of the grain, $ \bfo \cdot \bfn' = 0$. This leads to
\begin{equation}
0= -s_\alpha(s_\varphi c_\gamma+ c_\varphi c_\theta s_\gamma) + c_\alpha s_\theta s_\gamma
\end{equation}
which can be solved for $\gamma$,
\begin{equation}
\gamma=\tan^{-1}\left(\frac{s_\alpha s_\varphi}{c_\alpha s_\theta - s_\alpha c_\varphi c_\theta}\right)
\end{equation}
The arc-tangent function is undefined for $0/0$. This happens if
either $s_\varphi = 0$ and $\alpha = \theta$, or if $s_\alpha = 0$ and
$s_\theta=0$. The first case means that the scattering plane is the
plane of incidence (because $\varphi = 0$), therefore the plane of
scattering is the plane of departure as well, $\gamma = 0$ (or $\gamma
= \pi$). In the second case, the propagation direction before
scattering is (anti-)parallel to the grain orientation and the
scattering plane will be the plane of departure, again $\gamma = 0$.

\begin{table*}[h!tb]
\begin{center}
  \caption {Notation} \label{sec.notation}
 \begin{tabular}{|l|l|}
\hline
\hline
    Symbol & Description \\
  \hline
    $i,j,m$ & numbering indices \\
    $(x,y,z)$ & axis of Cartesian system \\
    $\alpha$ & angle of incidence, Fig.~\ref{p2:fig:scattering_geometry} \\
  $\beta$ & rotation angle, esp. between North direction and the plane of incidence, $\beta := \sphericalangle{(\bfd_\E,\bfn)}$  \\
    $\gamma$ & exit angle, Fig.~\ref{p2:fig:scattering_geometry} \\
   $(\theta$\,, $\varphi)$ & pair of scattering angles  \\
    $(\theta_\text{obs}$\,, $\varphi_\text{obs})$ & pair of angles to observer \\
$N_\theta$, $N_\varphi$ & number of bins along scattering angles \\
 $\zeta $  &    random number \\
   $v_{ceil}$ & ``ceiling'' value for rejection sampling (maximum probability) \\
   $P(\theta, \varphi)$ & probability density function \\
   $s$, $s'$  & length of photon path through medium \\
  $\Delta s_j$ & length of incremental photon path in cube $j$ \\
      $\Delta$ & $10^4$ times the difference of $V/I$ of {\name} minus the analytical solution by \citetalias{Peest2017} \\
  $\bfr$ & position vector in the medium \\
  $\bfo$ & symmetry axis of grain along major axis, Fig.~\ref{p2:fig:scattering_geometry}  \\
    $\bfk$ & incoming photon direction, Fig.~\ref{p2:fig:scattering_geometry} \\
  $\bfk'$ & outgoing  photon direction \\
  $\bfk_\text{obs}$ & photon direction towards observer \\
    $\bfd_\N$ & North direction in the particle frame (plane of incidence)\\
    $\bfd_{\N,init}$ & initial North direction, Fig.~\ref{p2:fig:scattering_geometry} \\
    $\bfd_{\N,sca}$ & North direction in the plane of scattering before the scattering event, Fig.~\ref{p2:fig:scattering_geometry} \\
    $\bfd'_{\N}$ &  North direction in particle frame (plane of departure), Fig.~\ref{p2:fig:scattering_geometry} \\
    $\bfd'_{\N,sca}$ & North direction in the plane of scattering, outgoing after scattering process, Fig.~\ref{p2:fig:scattering_geometry} \\
      $\bfd_\E$ & East direction \\
   $\bfn$ & normal to the plane of incidence \\
   $\bfn'$ & normal to the plane of departure \\
  $I$ & intensity \\
  $I'$ & scattered intensity \\
$j$ & anisotropic emissivity\\
  $\Phi$ & scattering phase function \\
  $\cext$ & extinction cross section  \\
    $\csca$ & scattering cross section  \\
    $\cpol$ & dichroism cross section \\
    $\ccpol$ & circular polarization cross section \\
    $\cscapol$ & polarization scattering cross section \\
    $\Lambda$ & albedo \\
    $n$ &  density \\
    $\tau$ & optical depth \\
 $\tau_\text{out}$ & optical depth from present position to outer cloud boundary \\
    $\bfS$ & Stokes vector \\
    $(I,Q,U,V)$ & components of Stokes vector $\bfS$\\
    $\bfM$ & $4\times4$ M\"uller matrices \\
    $\bfR$  & rotation matrix \\
     $\bfR_{\text{obs}}$ & rotation from cell to observer frame\\
     $\bfZ(\bfk,\bfk')$ & scattering matrix \\
     $\bfK$ & extinction matrix \\
    $\bfE_j$ & dichroism and birefringence matrix \\
\hline
 \end{tabular}
 \end{center}
 \end{table*}


\end{document}